\documentclass[twocolumn,showpacs,amsmath,amssymb]{revtex4}
\usepackage[utf8]{inputenc}
\usepackage{bm}
\usepackage{amsmath}
\usepackage{graphicx}
\usepackage{esint}

\makeatletter

\DeclareTextSymbolDefault{\textquotedbl}{T1}

\@ifundefined{textcolor}{}
{%
 \definecolor{BLACK}{gray}{0}
 \definecolor{WHITE}{gray}{1}
 \definecolor{RED}{rgb}{1,0,0}
 \definecolor{GREEN}{rgb}{0,1,0}
 \definecolor{BLUE}{rgb}{0,0,1}
 \definecolor{CYAN}{cmyk}{1,0,0,0}
 \definecolor{MAGENTA}{cmyk}{0,1,0,0}
 \definecolor{YELLOW}{cmyk}{0,0,1,0}
}

\makeatother

\begin{document}
\title{Fast Functionalization with High Performance in the Autonomous Information
Engine }
\author{Zhiyu Cao}
\thanks{These authors have contributed equally to this work. }
\author{Ruicheng Bao}
\thanks{These authors have contributed equally to this work. }
\author{Jiming Zheng}
\author{Zhonghuai Hou}
\thanks{E-mail: hzhlj@ustc.edu.cn}
\affiliation{Department of Chemical Physics \& Hefei National Research Center for
Physical Sciences at the Microscale, University of Science and Technology
of China, Hefei, Anhui 230026, China }
\date{\today}
\begin{abstract}
Mandal and Jarzynski have proposed a fully autonomous information heat
engine, consisting of a demon, a mass and a memory register interacting
with a thermal reservoir {[}Proc. Natl. Acad. Sci. U.S.A. 109, 11641
(2012){]}. This device converts thermal energy into mechanical work
by writing information to a memory register, or conversely, erasing
information by consuming mechanical work. Here, we derive a speed
limit inequality between the relaxation time of state transformation
and the distance between the initial and final distributions, where
the combination of the dynamical activity and entropy production plays
an important role. Such inequality provides a hint that a speed-performance
trade-off relation exists between the relaxation time to functional
state and the average production. To obtain fast functionalization
while maintaining the performance, we show that the relaxation dynamics
of information heat engine can be accelerated significantly by devising
an optimal initial state of the demon. Our design principle is inspired
by the so-called Mpemba effect, where water freezes faster when initially
heated.
\end{abstract}
\maketitle
\textit{Introduction.}---Maxwell's demon is a device that can measure
the microstate of a closed system, thereby reducing its entropy, seemingly
in violation with the second law \cite{maxwell2001theory}. Discussions
on this thought experiment raged through most of the twentieth century
\cite{smoluchowski1927experimentell,szilard1929entropieverminderung,brillouin1951maxwell,penrose2005foundations,feynman2011feynman}
and found a full resolution with the works of Landauer and Bennett
\cite{landauer1961irreversibility,bennett1982thermodynamics}. Crucial
point to understand the problem is the fact that the demon needs to
store information about the gas particles and this involves increasing
the information entropy of the memory registers. Later, deleting this
information requires an increase in entropy such that the second law
is restored \cite{maruyama2009colloquium}. Due to experimental advances,
it is nowadays possible to control systems down to the nanoscale,
making it possible to realize Maxwell's demon in the laboratory \cite{serreli2007molecular,berut2012experimental,toyabe2010experimental,koski2014exper,koski2014experimental,koski2015chip,vidrighin2016photonic,cottet2017observing,kumar2018sorting,masuyama2018information,ribezzi2019large,paneru2020efficiency}.
Maxwell's \textquotedbl intelligent\textquotedbl{} demon provides an
excellent arena for studying the thermodynamic framework of information
processing \cite{zurek1989thermodynamic,maruyama2009colloquium,hosoya2011maxwell}.
Theoretical research on models fall into two categories: autonomous
\cite{mandal2012work,mandal2013maxwell,barato2014unifying,horowitz2014thermodynamics,strasberg2017quantum,joseph2021efficiency,lu2019programmable}
and feedback control loops \cite{barato2014unifying,horowitz2014thermodynamics,strasberg2017quantum,quan2006maxwell,abreu2011extracting,ribezzi2019large}.

Recently, physicists have devised a series of \textsl{autonomous}
models without the participation of any \textquotedbl intelligent\textquotedbl{}
demon, which can achieve Maxwell's original vision and obtain results
that are consistent with the predictions of Landauer's principle \cite{mandal2012work,mandal2013maxwell,barato2014unifying,horowitz2014thermodynamics,strasberg2017quantum,joseph2021efficiency,lu2019programmable}.
The construction of autonomous information heat engines or erasers
not only helps to understand the basic concepts of information thermodynamics,
but also has important application prospects. Particularly, Mandal
and Jarzynski proposed a fully autonomous information heat engine
(IHE) setup \cite{mandal2012work}. After a certain relaxation time
to enter the functional state, the IHE converts thermal energy into
mechanical work by writing information to a memory register, rectifying
thermal fluctuations, or conversely, by consuming mechanical work
to (partially) erase the information on the memory register. Going
a step further, they also considered an autonomous information refrigerator
(IR) model \cite{mandal2013maxwell}. Similar to the IHE model, the
refrigerator utilizes thermal fluctuations to transfer heat from a
low temperature thermal reservoir to a high temperature, or acts as
an eraser to reduce the information on the memory register after a
finite relaxation interval to function. The performance of the autonomous
IHE/IR is measured by the average production, which quantifies the
rectification of the bits. A natural thought is that one may expect
that an efficient model can quickly enter the functional state with
high-performance. Therefore, how to design the IHE to achieve fast
functionalization while maintaining high production is of great importance.

In this letter, we analyze both the speed (relaxation time to functional
state) and the performance (average production) of autonomous IHE
model. An inequality related the rate of the state transformation
to the distance between two probability distributions has been derived
to claim that there is a speed-performance trade-off relation between
the relaxation time and the production, highlighting that the IHE
cannot be functionalized quickly with high production for fixed entropy
production. To overcome this limitation, we are committed to develop
a design strategy that allows the IHE to move quickly into an highly
efficient state. Remarkably, we show that the IHE can always reach
the stationary functional state at a remarkable faster pace by specially
preparing the demon's initial state, which is reminiscent of the Markovian
Mpemba effect \cite{aristotle1933metaphysics,mpemba1969cool,lu2017nonequilibrium,klich2019mpemba,gal2020precooling,carollo2021exponentially,kumar2020exponentially,lasanta2017hotter,baity2019mpemba,gijon2019paths,torrente2019large,chetrite2021metastable,vadakkayil2021should,yang2020non,busiello2021inducing,schwarzendahl2021anomalous,santos2020mpemba,biswas2022mpemba}.

\begin{figure}
\begin{centering}
\includegraphics[width=0.8\columnwidth]{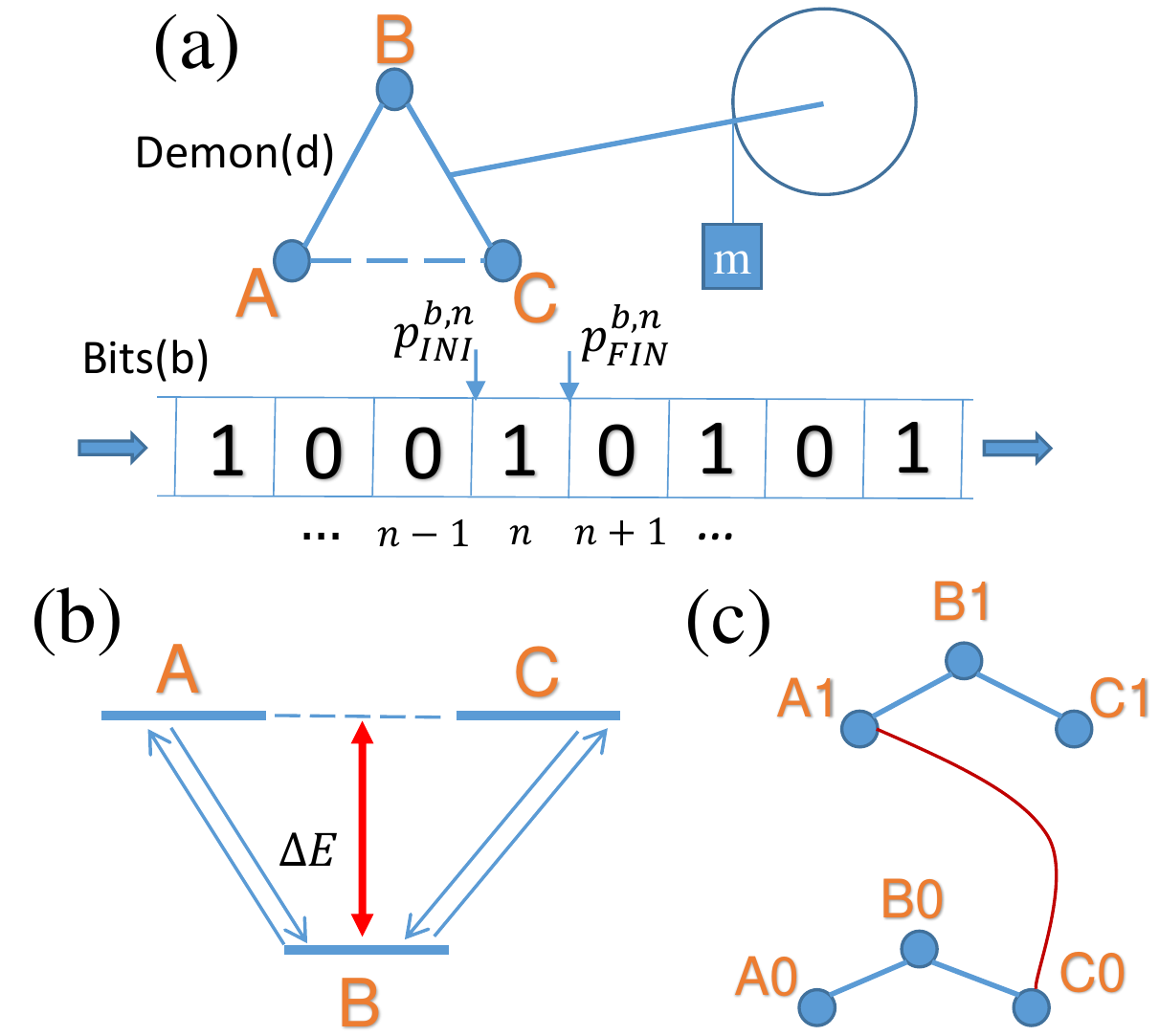}
\par\end{centering}
\caption{The information heat engine (IHE) setup. (a) The three-state demon
interacts with a sequence of bits, a mass and a reservoir. (b) Schematic
depiction of the demon. The state of the demon is indicated by an
arrow pointing in one of the directions. The states $B$ and $A/C$
are characterized by an energy difference $\Delta E=E_{u}-E_{d}$
with $E_{B}=E_{u}$ and $E_{A}=E_{C}=E_{d}$. (c) Network depiction
of the composite $6$-state system, showing allowed transitions. The
edge connecting $A_{1}$ and $C_{0}$ originates from the coupling
between demons and bits.}

\label{fig:1}
\end{figure}

\textit{Model.}---We start by introducing a modified IHE setup proposed
by Mandal and Jarzynski, as illustrated in Fig. \ref{fig:1}. In general,
the IHE model has $k$-state demon that interacts with a mass, a thermal
reservoir with temperature $T$, and a stream of bits (labeled $0$
and $1$), which acts as memory registers. The demon is initially
set up in contact with another thermal reservoir with temperature
$T_{\text{ini}}$ to reach equilibrium, and then it is coupled to
the memory registers, constituting a $2k$-states composite system.
At any instant in time, the bit stream moves through the demon in
a given sequence written in advance at a constant speed. After interacting
with the demon for a fixed time interval $\tau$, the bit moves forward,
and a new bit comes in. The demon transfers randomly between $k$
states, and the bit can transfer together with the demon from state
0 (1) to 1 (0). Net differences between the clockwise (CW) and counterclockwise
(CCW) cyclic transition will cause the demon to display directional
rotation and lift the mass.

Take the 3-state model ($k=3$, labeled $A$, $B$ and $C$) for example
shown in Fig. \ref{fig:1} (a), where the states $B$ and $A/C$ are
characterized by an energy difference $\Delta E=E_{u}-E_{d}$ with
$E_{B}=E_{u}$ and $E_{A}=E_{C}=E_{d}$ {[}c.f. Fig. \ref{fig:1}
(b){]}. It can be defined that the transition in the $A\rightarrow B\rightarrow C\rightarrow A$
direction is CW, and the transition in the opposite direction is CCW.
If the demon is uncoupled to the bit, it can only jump between states
$A$ and $B$, and $B$ and $C$, which means there is no net cycle
and the mass can not be lifted. However, if the demon interacts with
the bit, they will together form a composite system with six states
($A_{0}$, $A_{1}$, $B_{0}$, $B_{1}$, $C_{0}$, $C_{1}$). Thereupon,
the demon is allowed to transfer from $C$ to $A$ if the bit flips
from $0$ to $1$ simultaneously, and vice versa {[}c.f. Fig. \ref{fig:1}(c){]},
which means that the net flux of cycle can emerge due to such cooperative
transitions with the help of the bit stream. The average number of
net cycles can be identified as the production, which is the key performance
of IHE. The incoming bit stream of the IHE contains a mixture of $0$’s
and $1$’s with fixed probabilities $p_{0}$ and $p_{1}$, which are
statistically independent respectively. The evolution of the probability
distribution in every interval can be separated into two stages: (i)
the Markovian evolution of composite $2k$-state distribution governed
by transition matrix $\{R_{ij}\}$, and (ii) the projection process
at the end of every interval, which eliminates the correlations between
the $k$-state system and the bit. For a finite interval, the IHE
will relax to a periodic steady state for a large enough interval
number. Let $\delta=p_{0}-p_{1}$ denotes the proportional excess
of $0$’s among incoming bits initially. Once the demon has reached
its periodic steady state, let $p_{0}^{\prime}$ and $p_{1}^{\prime}$
denote the fractions of $0$’s and $1$’s in the outgoing bit stream,
and $\delta^{\prime}=p_{0}^{\prime}-p_{1}^{\prime}$. The circulation
as a measure of the average production of $1$’s per interaction interval
in the outgoing bit stream can then be defined as

\begin{equation}
\Phi=p_{1}^{\prime}-p_{1}=\frac{\delta-\delta^{\prime}}{2}.\label{eq:ap}
\end{equation}

Further elaborating the evolution of composite $2k$-state system,
we introduce the transition matrix $\mathcal{\bm{\mathcal{T}}=\mathcal{\bm{\mathcal{T}}}}_{k\times k}$
whose element $T_{\mu\nu}$ represents the probability for the demon
to be in state $\mu$ at the end of an interaction interval, given
that it was in state $\nu$ at the start of the interval. Let $\bm{p}_{\text{INI}}^{d,n}/\bm{p}_{\text{FIN}}^{d,n}$
($\bm{p}_{\text{INI}}^{b,n}/\bm{p}_{\text{FIN}}^{b,n}$) denote the
distribution of the demon (bit) at the start/end of the $n$-th interaction
interval, and $\bm{p}_{\text{INI}}^{b,0}=(p_{0},p_{1})$. The evolution
of the demon over many intervals is given by repeated application
of the matrix $\mathcal{\bm{\mathcal{T}}}$ {[}SM{]}. Because $\mathcal{\bm{\mathcal{T}}}$
is a positive transition matrix, the demon evolves to a periodic steady
state,

\begin{equation}
\lim_{n\to\infty}\bm{p}_{\text{INI}}^{d,n}=\lim_{n\to\infty}\mathcal{\bm{\mathcal{T}}}^{n}\bm{p}_{\text{INI}}^{d,0}=\bm{p}_{\text{INI}}^{d,ps}.\label{eq:taups}
\end{equation}
The unique periodic steady state can be obtained by solving $\mathcal{\bm{\mathcal{T}}}\bm{p}_{\text{INI}}^{d,ps}=\bm{p}_{\text{INI}}^{d,ps}$,
which is just the functional state of the IHE which can produce anomalous
work stably. Meanwhile, the bit distribution at the end of the $n$-th
interaction interval, $\bm{p}_{\text{FIN}}^{b,n}$, also converges
to a periodic steady state as $\lim_{n\to\infty}\bm{p}_{\text{FIN}}^{b,n}=\bm{p}_{\text{FIN}}^{b,ps}=(p_{0}^{\prime},p_{1}^{\prime})$.
Nevertheless, compared to the evolution of demon distribution {[}Eq.
(\ref{eq:taups}){]}, the evolution of bit distribution cannot be
simply described by a propagator due to the bit reset operation at
the start of each interval. For the 3-state model introduced above
{[}c.f. Fig. \ref{fig:1}{]}, the exact expression of the periodic
steady state $\bm{p}_{\text{INI}}^{d,ps}$ and average production
$\Phi$ can be obtained by solving the evolution theoretically, which
can be found in the Supplemental Material {[}SM{]}.

\textit{Speed-performance trade-off relation.}--- As stated above,
the demon will go through a certain number of time intervals before
reaching the periodic steady state, and the relaxation time for the
demon to move from an initial state to the functional state is another
crucial feature besides the production. In the following, we turn
to analyze the relationship between the relaxation time and average
production. We use the $L^{1}$-norm to measure the statistical distance
of two probability distributions $\bm{p}$ and $\bm{q}$, i.e., the
total variation distance reads as $D(\bm{p},\bm{q})=\left\Vert \bm{p}-\bm{q}\right\Vert =\sum_{i}\left|p_{i}-q_{i}\right|$.
When the demon reaches the periodic steady state, the composite system
reaches the functional state simultaneously. We assume that there
exists a critical interval number $N_{c}=N_{c}(d)$ satisfying $D(\bm{p}_{\text{FIN}}^{b,N_{c}},\bm{p}_{\text{FIN}}^{b,\infty})\leq d$,
which is expected to be proportional to the relaxation time $\tau_{c}=N_{c}\tau$
when the cut-off parameter $d$ is sufficiently small. Thus, it is
important to analyze the distance between the bit distribution of
the $n$-th interval $\bm{p}_{\text{FIN}}^{b,n}$ and the steady one
$\bm{p}_{\text{FIN}}^{b,\infty}$ at the end of each period, i.e.,
$D(\bm{p}_{\text{FIN}}^{b,n},\bm{p}_{\text{FIN}}^{b,\infty})$. The
average production are connected to the distance between the initial
bit state $\bm{p}_{\text{INI}}^{b,0}=(p_{0},p_{1})$ and the final
periodic steady state $\bm{p}_{\text{FIN}}^{b,\infty}=(p_{0}^{\prime},p_{1}^{\prime})$
as $2\Phi=D(\bm{p}_{\text{INI}}^{b,0},\bm{p}_{\text{FIN}}^{b,\infty})$.
Then, we perform an approximation that $2\Phi=D(\bm{p}_{\text{INI}}^{b,0},\bm{p}_{\text{FIN}}^{b,\infty})\approx D(\bm{p}_{\text{FIN}}^{b,0},\bm{p}_{\text{FIN}}^{b,\infty})$,
assuming that the difference between the initial bit state and final
one of the first interval is small. This assumption is based on the
intuition that the change of the bit state in a single interval will
not be particularly large.

As mentioned above, the evolution of the bit is non-Markovian (can
not be described by a propagator), so it is difficult to explore the
convergence of the bit distribution $\bm{p}_{\text{FIN}}^{b,n}$,
i.e., $D(\bm{p}_{\text{FIN}}^{b,n},\bm{p}_{\text{FIN}}^{b,\infty})$.
However, the evolution of demon distribution is easier to capture,
so we develop an information-theoretical relation between the distance
function of the bit and demon to face this difficulty, which reads

{}
\begin{equation}
D(\bm{p}_{\text{FIN}}^{b,n},\bm{p}_{\text{FIN}}^{b,0})\le D(\bm{p}_{\text{INI}}^{d,n},\bm{p}_{\text{INI}}^{d,0})\label{eq:hie}
\end{equation}
for any interval number $n$ {[}SM{]}. Eq. (\ref{eq:hie}) shows that
the distance between final bit distribution of the $n$-th interval
and initial one is always smaller than the distance between initial
demon distribution of the $n$-th interval and initial one, which
serves as a hierarchy of the distance function between the distribution
of demon and bit. Physically, such hierarchical relation can be interpreted
as the bit distance is the projection of the demon distance in lower
dimensions, determined by the unusual dynamics of IHE {[}SM{]}. Based
on Eq. (\ref{eq:hie}), the relationship between the average production
and the demon distance can be obtained as
\begin{equation}
2\Phi\lesssim D(\bm{p}_{\text{INI}}^{d,0},\bm{p}_{\text{INI}}^{d,\text{\ensuremath{\infty}}}).
\end{equation}
Without loss of generality, we further assume the transition matrix
$\mathcal{\mathcal{\bm{\mathcal{T}}}}$satisfies the detailed balance
condition \cite{mandal2012work}. By using the hierarchical structure,
a speed limit inequality between the critical interval number $N_{c}$
(i.e., the relaxation time) and average production can be obtained
as $\Phi$

\begin{equation}
N_{c}\ge N_{\text{SL}}=\frac{\Phi}{\sqrt{\frac{1}{2}\text{\ensuremath{\dot{\Sigma}_{N_{c}}}}\cdot\langle A\rangle_{N_{c}}}}.\label{eq:SL}
\end{equation}
Here, $\Sigma_{N_{c}}\equiv\sum_{n=0}^{N_{c}}\Delta_{n}S$ is the
total entropy production with $\Delta_{n}S\equiv\sum_{i,j}\mathcal{T}_{ij}p_{\text{INI},j}^{d,n}\ln\left(\mathcal{T}_{ij}p_{\text{INI},j}^{d,n}/\mathcal{T}_{ji}p_{\text{INI},i}^{d,n}\right)$
the entropy production for the $n$th interaction interval, and $\dot{\Sigma}_{N_{c}}=\Sigma_{N_{c}}/N_{c}$
is its average rate. The dynamical activity $A(n)=\sum_{i\neq j}\mathcal{T}_{ij}p_{\text{INI},j}^{d,n}$
and its time average $\langle A\rangle_{N_{c}}=N_{c}^{-1}\sum_{n=0}^{N_{c}}A(n)$
quantify how frequently jumps between different states occur, i.e.,
the time scale of the system \cite{shiraishi2018speed,lecomte2007thermodynamic,garrahan2007dynamical,baiesi2009fluctuations,baiesi2009nonequilibrium,maes2017non,di2018kinetic}.
The novelty of this nontrivial relation reveals that there exists
a speed-performance trade-off between the relaxation time and average
production, highlighting that the IHE cannot be functionalized quickly
with high production for fixed entropy production. The structure of
Eq.(\ref{eq:SL}) also reminds us of the conventional quantum speed
limit \cite{mandelstam1991uncertainty,fleming1973unitarity,anandan1990geometry,margolus1998maximum,pfeifer1993fast,taddei2013quantum,del2013quantum,deffner2013quantum,pires2016generalized,funo2017universal,deffner2017geometric},
which is an important issue relevant to broad research fields including
quantum control theory and have been extended to the case of classical
dynamics recently \cite{shanahan2018quantum,okuyama2018quantum,ito2018stochastic,shiraishi2018speed,shiraishi2019information,nicholson2020time,ito2020stochastic,van2020unified,gupta2020tighter,yoshimura2021thermodynamic}.
In addition, we state that a series of similar inequalities can be
obtained and the result of Eq. (\ref{eq:SL}) can be further improved.
Among them, the tightest form reads

\begin{equation}
N_{c}\geq N_{TSL}=\frac{2\Phi}{\dot{\Sigma}_{N_{c}}}\cdot f\left(\frac{\dot{\Sigma}_{N_{c}}}{2\langle A\rangle_{N_{c}}}\right).\label{eq:tsl}
\end{equation}
Eqs. (\ref{eq:SL}) and (\ref{eq:tsl}) constitute our first important
result. Here, $f(x)$ is the inverse function of $x\tanh(x)$, and
the concavity property of the function $(x^{2}/y)f(x/y)^{-2}$ ensures
that Eq. (\ref{eq:tsl}) is tighter than Eq. (\ref{eq:SL}) with $N_{c}\geq N_{TSL}\ge\max\left\{ N_{\text{SL}},\Phi/\langle A\rangle_{N_{c}}\right\} $
\cite{vo2022unified,lee2022speed}. The detailed derivation has been
provided in the Supplemental Material (SM). Finally, we reiterate
that the speed limit holds for the detailed balance case and declare
that an analogous relation for the general case without detailed balance
condition can also be obtained in a similar way, where the excess
entropy production \cite{hatano2001steady} plays a substitute role
as the conventional total entropy production \cite{shiraishi2018speed}.

\begin{figure}
\begin{centering}
\includegraphics[width=0.8\columnwidth]{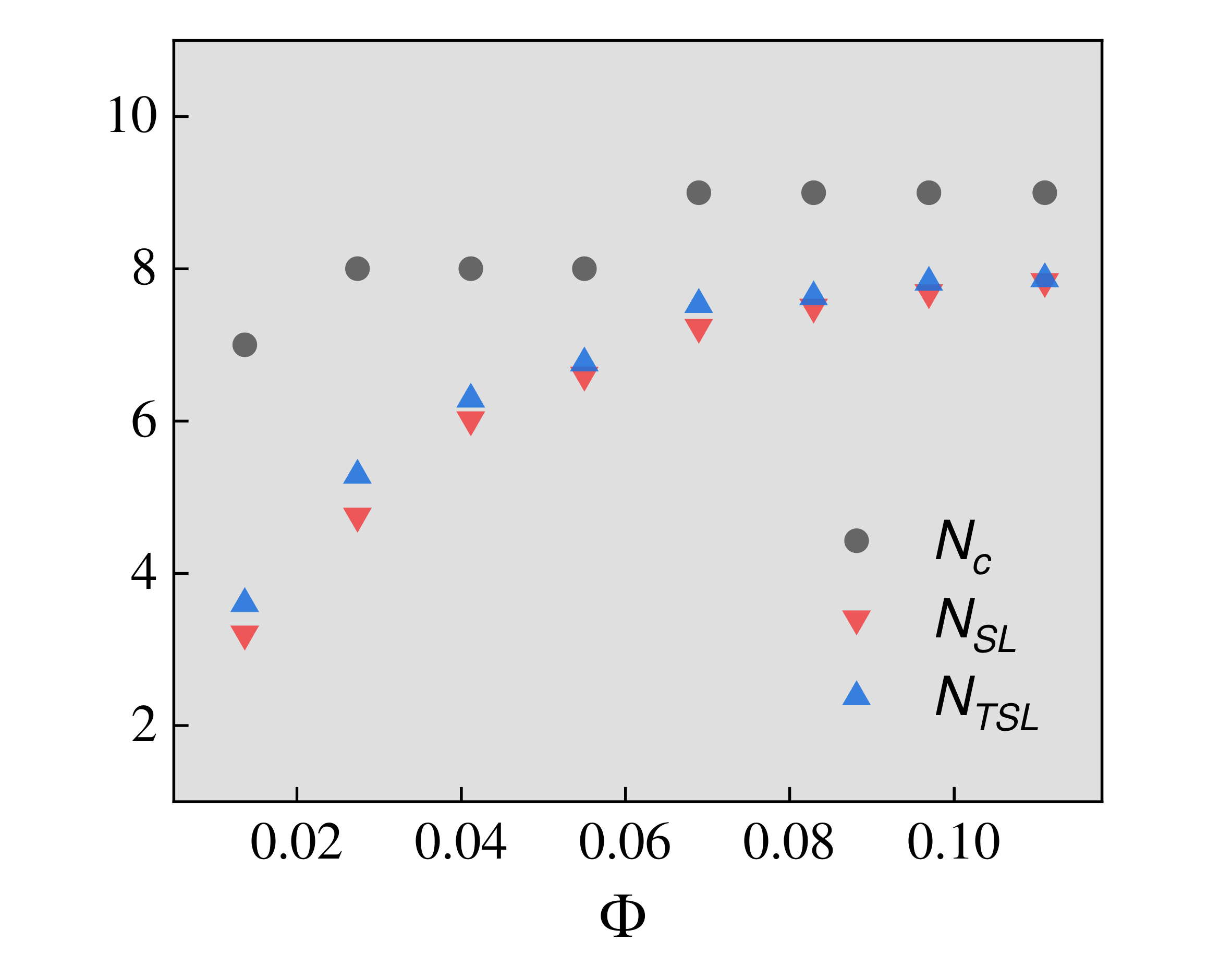}
\par\end{centering}
\caption{Demonstration of the speed-limit inequalities for the $3$-state model.
The critical interval number $N_{c}$ (black circles), speed limit
bound $N_{\text{SL}}$ (orange down triangles) and its tighter form
$N_{TSL}$ (blue up triangles) have been shown as functions of the
average production $\Phi$, where $f(x)$ is the inverse function
of $x\tanh(x)$. Both Eqs. (\ref{eq:SL}) and (\ref{eq:tsl}) have
been verified. Different points are obtained from varying $\text{\ensuremath{\delta}}$
with fixed $\tau=1.0$ and $\delta=0.1$. }

\label{fig:2}
\end{figure}

Here, we demonstrate the speed limit inequality with the $3$-state
IHE model introduced above, c.f. Fig. \ref{fig:1}. The control parameters
in this model are the weight parameter $\epsilon$, excess of the
incoming bit $\delta$ and the time interval $\tau$. More detailed
descriptions have been provided in the SM {[}SM{]}. By setting the
cut-off parameter $d=10^{-6}$, the critical interval number $N_{c}$,
entropy production rate $\dot{\Sigma}_{N_{c}}$ and the average dynamical
activity $\langle A\rangle_{N_{c}}$ can be obtained from numerically
operating the convergence of demon state, $\bm{p}_{\text{INI}}^{d,n}=\mathcal{\bm{\mathcal{T}}}^{n}\bm{p}_{\text{INI}}^{d,0}$.
Meanwhile, the average production $\Phi$ can be obtained from the
exact expression {[}SM{]}. We depict the critical interval number
$N_{c}$ (black circles), the speed limit bound $N_{SL}$ (orange
down triangles) and its tighter form $N_{TSL}$ (blue up triangles)
as functions of the average production $\Phi$ by varying $\delta$
with fixed $\tau=1.0$ and $\epsilon=0.1$ in Fig. \ref{fig:2}. Our
speed limits Eq.(\ref{eq:SL}) and Eq.(\ref{eq:tsl}) are valid, which
provides the hint of the speed-performance trade-off and do reasonable
job of predicting the critical functionalization interval of the IHE.

\textit{Fast functionalization.}---Due to the speed-performance trade-off
relation, we are motivated to find a design strategy to speed up the
functionalization while maintaining the average production. To this
goal, we analyze the relaxation modes and timescales based on the
framework of spectral decomposition. As mentioned above, the matrix
$\mathcal{\bm{\mathcal{T}}}$ governs the evolution of the system.
The eigenvalues of $\mathcal{\bm{\mathcal{T}}}$ is related to the
timescale for different dynamical modes, which can be numbered in
a descending order: $1=\mu_{1}>\left|\mu_{2}\right|\ge\left|\mu_{3}\right|\ge...$
(here, we assume that $\mu_{1}$ is not degenerate). Accordingly,
the right $\bm{R}_{i}$ and left eigenvectors $\bm{L}_{i}$ for an
eigenvalue $\mu_{i}$ correspond to the $i$-th dynamical modes, which
read $\mathcal{\mathcal{\bm{\mathcal{T}}}}\bm{R}_{i}=\mu_{i}\bm{R}_{i}$
and $\bm{L}_{i}^{\text{T}}\mathcal{\mathcal{\bm{\mathcal{T}}}}=\mu_{i}\bm{L}_{i}^{\text{T}}$,
respectively. The right eigenvector $\bm{R}_{1}$ for the principal
eigenvalue $1=\mu_{1}$ is the functional periodic steady state, $\bm{R}_{1}=\bm{p}_{\text{INI}}^{d,ps}$,
and its corresponding left eigenvector is the identity. The spectral
decomposition allows us to expand any probability distribution as
a linear combination of the eigenvectors. Particularly, for the initial
state $\bm{p}_{\text{INI}}^{d,0}$, it can be written as

\begin{equation}
\bm{p}_{\text{INI}}^{d,0}=\bm{p}_{\text{INI}}^{d,ps}+\sum_{i>1}d_{i}\bm{R}_{i},\label{eq:0th}
\end{equation}
where the corresponding overlap coefficient between the initial probability
and the $i$-th left eigenvector $\bm{L}_{i}^{\text{T}}$ is
\begin{equation}
d_{i}=\frac{\bm{L}_{i}^{\text{T}}\cdot\bm{p}_{\text{INI}}^{d,0}}{\bm{L}_{i}^{\text{T}}\cdot\bm{R}_{i}}.\label{eq:di}
\end{equation}
During the relaxation process, the initial demon distribution of the
$n$-th time interval, $\bm{p}_{\text{INI}}^{d,n}$, can then be obtained
as

\begin{equation}
\bm{p}_{\text{INI}}^{d,n}=\mathcal{\bm{\mathcal{T}}}^{n}\bm{p}_{\text{INI}}^{d,0}=\bm{p}_{\text{INI}}^{d,ps}+\sum_{i>1}d_{i}\mu_{i}^{n}\bm{R}_{i}.\label{eq:disn}
\end{equation}
We once again emphasize that the eigenvectors can be interpreted as
dynamical modes that transport probability density from one part of
the conformational space to another, and the modules of the eigenvalues
give the relaxation rates of all the modes which has been excited.
From Eq.(\ref{eq:disn}), we find that the second eigenvalue $\left|\mu_{2}\right|$
determines the spectral gap, characterizing the longest timescale
of the relaxation, and $\bm{R}_{2}$ is in fact the slowest decaying
mode of the demon. Hence, the probability distribution (\ref{eq:disn})
can be approximated after a long time as $\mathcal{\bm{\mathcal{T}}}^{n}\bm{p}_{\text{INI}}^{d,0}\approx\bm{p}_{\text{INI}}^{d,ps}+d_{2}\mu_{2}^{n}\bm{R}_{2}$.

The explicit expression of the distance between the initial demon
distribution of the $n$-th time interval and the functional state
can be measured by the $L^{1}$-norm, which reads $\left\Vert \mathcal{\bm{\mathcal{T}}}^{n}\bm{p}_{\text{INI}}^{d,0}-\bm{p}_{\text{INI}}^{d,ps}\right\Vert =\sum_{i>1}d_{i}\left|\mu_{i}\right|_{q}^{n}\left\Vert R_{i}\right\Vert $.
Since the second eigenvalue $\left|\mu_{2}\right|$ determines the
decaying process dominantly, the relaxation timescale $\tau_{c}\equiv N_{c}\tau$
can be typically characterized as $\tau_{c}\sim\tau_{2}=-1/\text{\ensuremath{\ln}}\left|\mu_{2}\right|$,
i.e., the critical interval number $N_{c}$ is proportional to the
relaxation timescale as $N_{c}\propto\tau_{2}$ for a non-degenerate
system. Further, it can be found that the decaying rate of process
depends mostly on the overlap between between the initial probability
and the dominant mode $d_{2}$. A smaller $d_{2}$ implies that the
dominant mode are excited moderately and will induce a faster relaxation.
Such mechanism is in spirit similar to some kind of anomalous relaxation
referred as the Markovian Mpemba effect \cite{lu2017nonequilibrium},
where initiating the system at a hot temperature results in faster
cooling down than any colder temperature when the system is coupled
to a cold bath. Generally, the dynamics of the demon overlaps with
all decaying modes, particularly the slowest one. However, it can
be observed that the slowest mode can be completely depopulated initially
if there exists an equilibrium initial demon distribution $\bm{\pi}_{\text{INI}}^{d}$
satisfying $d_{2}|_{\bm{p}_{\text{INI}}^{d,0}=\bm{\pi}_{\text{INI}}^{d}}=0$,
i.e.,
\begin{equation}
\bm{L}_{2}^{\text{T}}\cdot\bm{\pi}_{\text{INI}}^{d}=0,
\end{equation}
Reasonably, for the IHE, the specific initial state $\bm{\pi}_{\text{INI}}^{d}=\bm{\pi}_{\text{INI}}^{d}(T_{opt})$
can be obtained by preparing the demon in a thermal reservoir with
an optimal temperature $T_{\text{ini}}=T_{\text{opt}}\neq T$. To
be specific, by preparing the demon's initial state $\bm{p}_{\text{INI}}^{d,0}=\bm{\pi}_{\text{INI}}^{d}(T_{opt})$
orthogonal to $\boldsymbol{L}_{2}$, the state converges at a shorter
timescale $\tau_{3}=-1/\text{\ensuremath{\ln}}\left|\mu_{3}\right|$,
which is in favor of a remarkable faster pace of the relaxation. Particularly,
for the 3-state IHE with $\bm{L}_{2}=(L_{21},L_{22},L_{23})$, the
optimal temperature can be solved as {[}SM{]}

\begin{equation}
T_{\text{opt}}/\Delta E=\left[k_{b}\ln(-\frac{L_{21}+L_{23}}{L_{22}})\right]^{-1},\label{eq:opt}
\end{equation}
which is the second main result of our paper. Here, $k_{b}$ is the
Boltzmann constant. The basic mechanism underpinning such phenomenon
is reminiscent of the strong Mpemba effect (SME) \cite{klich2019mpemba},
which has been verified by experiments in colloidal systems \cite{kumar2020exponentially}.
The set of initial states whose projection along $\bm{L}_{2}^{\text{T}}$
vanishes will identify a ($k-2$)-manifold which is referred to as
the strong Mpemba space (SM space), suggesting that the SME can only
exist in the model with more than two states. Although the relationship
between the energy landscape and approach to stationary state is generally
complex and volatile, the SME shows that special initial state setups
will induce a shortcut in relaxation, providing a useful recipe for
the design of high-quality information machine.

\begin{figure}
\begin{centering}
\includegraphics[width=1\columnwidth]{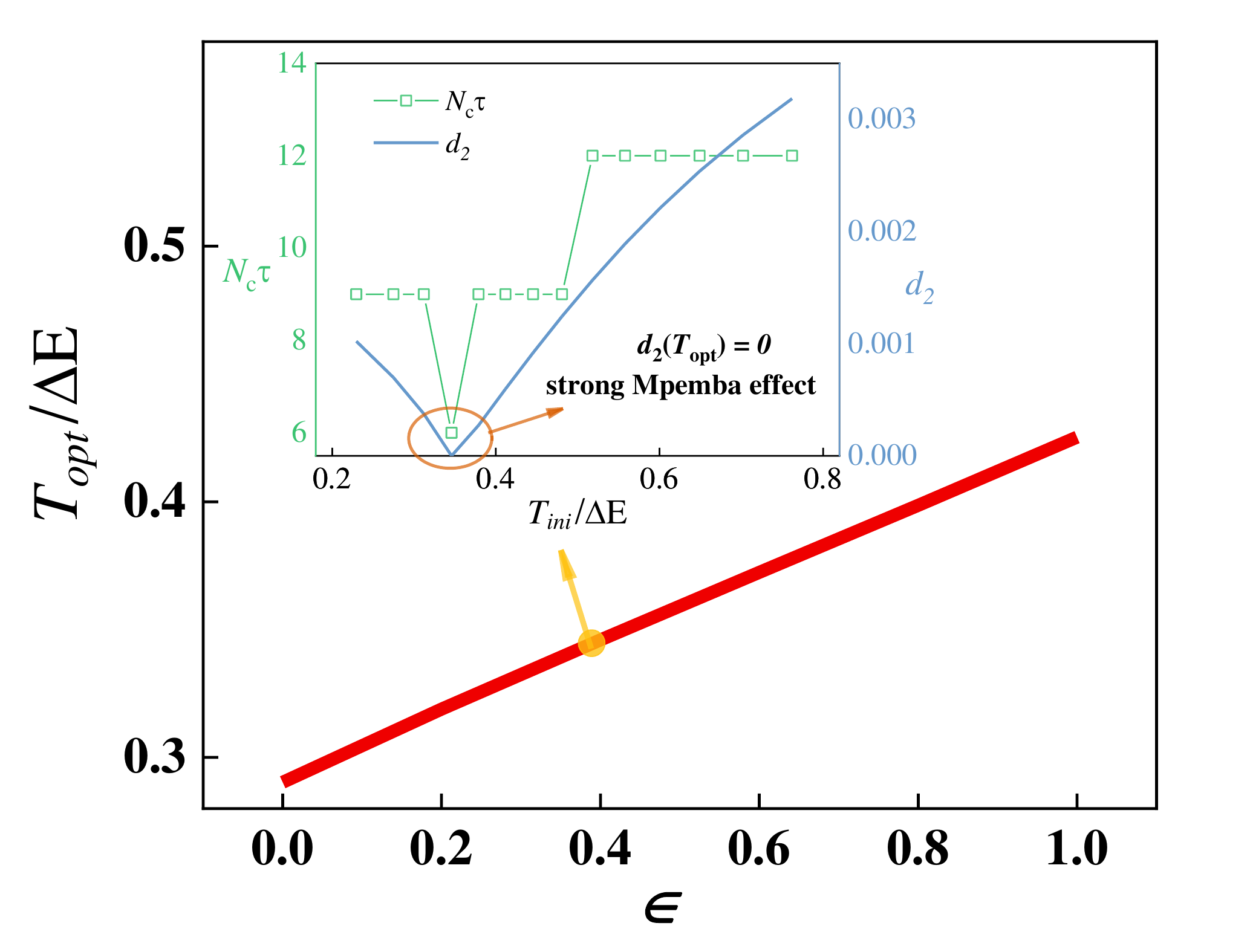}
\par\end{centering}
\caption{Demonstration of the strong Mpemba effect (SME). Optimal initial temperatures
of demon $T_{opt}/\Delta E$ {[}obtained from Eq. (\ref{eq:opt}){]}
are plotted as a function of the weighted parameter $\epsilon$ (the
red line). Parameter $\tau=3.0$ and $\delta=0.5$ are fixed. Inset
shows the critical relaxation time $N_{c}\tau$ (green squares) and
the overlap between between the initial probability and the dominant
mode $d_{2}$ (the blue line) as a function of the initial demon temperature
$T_{ini}/\Delta E$ for $\epsilon=0.4$. When $d_{2}(T_{opt})=0$,
the relaxation time reaches its minimum and the SME occurs.}

\label{fig:3}
\end{figure}

In the following, we illustrate that the 3-state IHE model allows
us to demonstrate how the SME controls the timescale for the approach
to the functional state. In Fig. \ref{fig:3}, the optimal initial
temperatures of demon $T_{opt}/\Delta E$, which can be theoretically
predicted from Eq. (\ref{eq:opt}), have been plotted as a function
of the weighted parameter $\epsilon$ (the red line) with $\tau=3.0$
and $\delta=0.5$. Also, the critical relaxation time $N_{c}\tau$
(green squares) can be numerically calculated, and the overlap between
between the initial probability and the dominant mode $d_{2}$ (the
blue line) can be obtained from the theoretical result of the spectral
decomposition of $\mathcal{\bm{\mathcal{T}}}$ {[}SM{]}. Both $N_{c}\tau$
and $d_{2}$ have been depicted as a function of the initial demon
temperature $T_{ini}/\Delta E$ for fixed $\epsilon=0.4$ in the inset
of Fig. \ref{fig:3}. Remarkably, it can be found that the relaxation
process is typically accelerated for the certain initial demon temperature
where the SME occurs with $d_{2}(T_{opt})=0$, demonstrating the applicability
of our design principle.

\textit{Discussion.}---In summary, we have derived speed limit inequalities
for the case of information machine. These results can be useful for
understanding trade-off relation between the speed of relaxation to
the functional state and the average production. To address this issue,
we have presented a design principle to shorten the timescale for
the approach to the functional state. Without sacrificing the production,
speed of relaxation can be accelerated by rationally designing the
initial state of the demon so that the slowest relaxation mode is
no longer excited. We have demonstrated our results by numerical verification.

Besides the speed limit we derived, other techniques might also be
hopeful to reveal such relation, such as the thermodynamic uncertainty
relation for arbitrary initial states \cite{liu2020thermodynamic}
and the information geometry \cite{ito2018stochastic,ito2020stochastic}.
And, the trade-off between speed and performance may be widespread
in other systems, such as periodic heat engine. Since the existence
of SME is robust to perturbations and can be observed even in the
thermodynamic limit \cite{klich2019mpemba}, we believe that the proposed
design principle is readily accessible in experiments. Finally, we
also suggest another interesting avenue deserving special attention
is that the stochastic resetting mechanism can further improve the
speed and performance of information engine \cite{bao2022designing}.
\begin{acknowledgments}
This work is supported by MOST(2018YFA0208702), NSFC (32090044, 21790350).
\end{acknowledgments}

\appendix
\onecolumngrid

\section*{Supplementary Material for ``Faster Functionalization with High
Performance in the Autonomous Information Engine''}


\section{Details of the model}

Here, we introduce the modified IHE model in more details, as shown
in Fig. \ref{fig:1}. As mentioned in the main text, the IHE model
has $k$-state (here, $k=3$) demon that interacts with: a thermal
reservoir, a mass that can be lifted or lowered, and a stream of bits
(labeled $0$ and $1$). When the bit moves forward, the demon transitions
between the $A$, $B$ and $C$ states simultaneously. When uncoupled
to the bit, the demon can jump between states $A$ and $B$, and $B$
and $C$. With the help of the bit stream, the demon and bit together
form a composite system with six states, $A_{0},\dots,C_{1}$, which
allows the transitions between $A$ and $C$ for the demon. Precisely,
when the demon interacts with the bit, the demon can transition from
$C$ to $A$ if the bit flips from $0$ to $1$ simultaneously, and
vice versa, as shown in Fig. \ref{fig:1}(c). The frequency difference
between the CW transition and the CCW transition will cause the demon
to display directional rotation.

We consider a positive positive external load $f=mg\Delta h/k_{b}T>0$
($T$ is the temperature of the thermal reservoir and $k_{b}$ is
Boltzmann constant), assuming that the mass $m$ is lifted by $\Delta h$
every time the demon makes a transition $C\rightarrow A$, and lowered
with $A\rightarrow C$. The transition rates with detailed balance
can be written as
\begin{equation}
\frac{R_{A,B}}{R_{B,A}}=\frac{R_{C,B}}{R_{B,C}}=e^{-\Delta E/k_{b}T},
\end{equation}
and
\begin{equation}
\frac{R_{A_{1},C_{0}}}{R_{C_{0},A_{1}}}=e^{-f}.
\end{equation}
For convenience, we set the $\Delta E\ll T\sim mg\Delta h$ so that
$R_{ij}=1$ for all transition rates except $R_{A_{1},C_{0}}$ and
$R_{C_{0},A_{1}}$. In particular, when the demon interacts with a
fixed bit for a long enough time, both of them will reach equilibrium
simultaneously, whose distribution read
\begin{equation}
p_{i,\:i\in\{A_{0},B_{0},C_{0}\}}^{eq}=\frac{e^{f}}{Z},\qquad p_{i,\:i\in\{A_{1},B_{1},C_{1}\}}^{eq}=\frac{1}{Z}
\end{equation}
with $Z=3(1+e^{f})$. For simplicity, a weight parameter $\epsilon$
is defined to describe the difference between the equilibrium probabilities
for the bit after summing over the states of the demon,

\begin{equation}
\epsilon\equiv p_{0}^{eq}-p_{1}^{eq}=\tanh(\frac{f}{2}).
\end{equation}
We assume that the incoming bit stream contains a mixture of $0$’s
and $1$’s, with probabilities $p_{0}$ and $p_{1}$, respectively,
with no correlations between bits. As stated in the main text, $\delta=p_{0}-p_{1}$
denotes the proportional excess of 0’s among incoming bits. The evolution
of the composite six-state system in every interval can be separated
into two stages: (i) the dynamic evolution governed by transition
rates $\{R_{ij}\}$, and (ii) the projection process at the end of
every interval, which eliminates the correlations between the three-state
system and the bit. For a finite $\tau$, the IHE will reach a periodic
steady state for large enough period number. Once the demon has reached
its periodic steady state, let $p_{0}$ and $p_{1}$ denote the fractions
of $0$’s and $1$’s in the outgoing bit stream, and let $\delta^{\prime}=p_{0}^{\prime}-p_{1}^{\prime}$
denote the excess of outgoing $0$’s. The circulation as a measure
of the average production of $1$’s per interaction interval in the
outgoing bit stream, which is the key performance of the engine, can
is defined as $\Phi=p_{1}^{\prime}-p_{1}=\left(\delta^{\prime}-\delta\right)/2$
{[}c.f. Eq. (\ref{eq:ap}) in the main text{]}.

\section{Derivation of the speed limit}

\subsection{Solving for the average production}

Firstly, we show the exact expression of average production, which
can be obtained by solving the periodic steady state as \cite{mandal2012work}

\begin{equation}
\Phi(\delta,\epsilon;\tau)=\frac{\delta-\epsilon}{2}\left[1-K(\tau)+\frac{\tau\delta}{6}J(\tau,\epsilon\delta)\right]\label{eq:phi}
\end{equation}
with

\begin{equation}
K(\tau)=e^{-2\tau}\frac{\left(1+8\alpha+4\sqrt{3}\beta\right)-\left(2+7\alpha+4\sqrt{3}\beta\right)e^{-2\tau}}{3-\left(2+\alpha\right)e^{-2\tau}}
\end{equation}
and

\begin{equation}
J(\tau,\epsilon\delta)=\frac{\left(1-e^{-\tau}\right)\left[2e^{-2\tau}\left(\alpha+\sqrt{3}\beta-1\right)\right]^{2}}{\left[3\left(1-\epsilon\delta e^{-\tau}\right)-\left(1-\epsilon\delta\right)\left(2+\alpha\right)e^{-2\tau}\right]\left[3-\left(2+\alpha\right)e^{-2\tau}\right]}.
\end{equation}
Here, $\alpha=\cosh(\sqrt{3}\tau)$, $\beta=\sinh(\sqrt{3}\tau)$.
Then, we provide key steps in the derivation of the periodic steady
state. We will use the notation $\bm{p}^{d}=\{p_{i}^{d}\}$ to denote
the probability distribution of the demon, $\bm{p}^{b}=\{p_{i}^{b}\}$
to denote the distribution of the bit. For the IHE model, $\bm{p}^{d}=(p_{A},p_{B},p_{C})^{\text{T}}$,
$\bm{p}^{b}=(p_{0},p_{1})^{\text{T}}$, and $\bm{p}=(p_{A0},p_{B0},p_{C0},p_{A1},p_{B1},p_{C1})^{\text{T}}$
is the joint probability distribution. The transition matrix $\mathcal{\bm{\mathcal{T}}=\mathcal{\bm{\mathcal{T}}}}_{3\times3}$
whose element $T_{\mu\nu}$ ($\mu$, $\nu$$\in\{A,B,C\}$) represents
the probability for the demon to be in state $\mu$ at the end of
an interaction interval, given that it was in state $\nu$ at the
start of the interval. Let $\bm{p}_{0}^{d}$ denote the distribution
of the demon at the start of a given interaction interval. The evolution
of the demon over many intervals is given by repeated application
of the matrix $\mathcal{\bm{\mathcal{T}}}$, which can be written
as

\begin{equation}
\bm{\mathcal{T}}=\mathcal{P}^{d}e^{\mathcal{R}\tau}\mathcal{M}.\label{eq:tm}
\end{equation}
Here, $\mathcal{P}^{d}=\left(\mathbb{I},\mathbb{I}\right)$ and $\mathcal{M}=\begin{pmatrix}p_{0}\mathbb{I}\\
p_{1}\mathbb{I}
\end{pmatrix}$ with $\mathbb{I}$ the identity matrix. $\mathcal{P}^{d}$ projects
out the state of the bit and $\mathcal{M}\bm{p}_{\text{INI}}^{d}$
gives the composite state of the initially uncorrelated demon and
bit. The transition rate matrix for the demon and the interacting
bit reads

\begin{equation}
\mathcal{R}=\begin{pmatrix}-1 & 1 & 0 & 0 & 0 & 0\\
1 & -2 & 1 & 0 & 0 & 0\\
0 & 1 & -2+\epsilon & 1+\epsilon & 0 & 0\\
0 & 0 & 1-\epsilon & -2-\epsilon & 1 & 0\\
0 & 0 & 0 & -2 & -2 & 1\\
0 & 0 & 0 & 1 & 1 & -1
\end{pmatrix},
\end{equation}
whose diagonal elements are determined by the requirement that the
elements in each column sum to zero. This matrix has six real, non-degenerate
eigenvalues that are (surprisingly) independent of $\epsilon$:

\begin{equation}
\{\lambda_{i}\}=\{0,-c,-1,-2,-3,-d\},
\end{equation}
where $a=1-\sqrt{3}$, $c=2-\sqrt{3}$, $x=1+\epsilon,$$b=1+\sqrt{3}$,
$d=2+\sqrt{3}$ and $y=1-\epsilon$. The quantities $a$, $b$, $x$
and $y$ will be used momentarily. As discussed in the main text,
the demon evolves to a periodic steady state,

\begin{equation}
\lim_{n\to\infty}\mathcal{\bm{\mathcal{T}}}^{n}\bm{p}_{\text{INI}}^{d,0}=\bm{p}_{\text{INI}}^{d,ps}\label{eq:taups-1}
\end{equation}
with $\mathcal{\bm{\mathcal{T}}}\bm{p}_{\text{INI}}^{d,ps}=\bm{p}_{\text{INI}}^{d,ps}$
gives the marginal distribution of the demon at the start of each
interaction interval.

The existence and uniqueness is guaranteed by the Perron-Frobenius
theorem. After a straightforward calculation we obtain that

\begin{equation}
\mathcal{\bm{\mathcal{T}}}=\frac{1}{12}\begin{pmatrix}F+G+\delta H & M-2\delta L & F-G+\delta H\\
M & M+12\sigma^{3} & M\\
F-G-\delta H & M+2\delta L & F+G-\delta H
\end{pmatrix}+\frac{\epsilon}{12}\begin{pmatrix}F-G-H & M+2L & F-G-H\\
0 & 0 & 0\\
-F+G+H & -M-2L & -F+G+H
\end{pmatrix},
\end{equation}
where $\sigma=e^{-\tau}$, $F=4+2\sigma^{3}$, $G=4\sigma^{2}+\sigma^{c}+\sigma^{d}$,
$H=\sqrt{3}(\sigma^{c}-\sigma^{d}),$$L=2\sigma^{2}-\sigma^{c}-\sigma^{d}$,
and $M=4-4\sigma^{3}$. By solving the equation $\mathcal{\bm{\mathcal{T}}}\bm{p}_{ini}^{d,ps}=\bm{p}_{ini}^{d,ps}$,
the periodic steady state reads
\begin{equation}
\bm{p}_{\text{INI}}^{d,ps}=\frac{1}{3}\begin{pmatrix}1+N\\
1\\
1-N
\end{pmatrix},\qquad N(\delta,\epsilon)=\frac{(\delta-\epsilon)(H-L)}{6-G+\delta\epsilon(G-6\sigma)}.
\end{equation}

To understand Eq.(\ref{eq:tm}), let $\bm{p}_{\text{INI}}^{d}$ denote
the distribution of the demon at the start of a given interaction
interval. $\bm{p}_{\text{INI}}=\mathcal{M}\bm{p}_{\text{INI}}^{d}$
gives the initial joint distribution of the demon and the incoming
bit. From this initial distribution, the joint state evolves under
the master equation $d\bm{p}/dt=\mathcal{R}\bm{p}$, then $\bm{p}_{\text{FIN}}=e^{\mathcal{R}\tau}\mathcal{M}\bm{p}_{\text{INI}}^{d}$
gives the joint distribution at the end of the interaction interval.
The matrix $\mathcal{P}^{d}$ then projects out the state of the bit,
thus $\bm{p}_{\text{FIN}}^{d}=\mathcal{P}^{d}e^{\mathcal{R}\tau}\mathcal{M}\bm{p}_{\text{INI}}^{d}=\bm{\mathcal{T}}\bm{p}_{\text{INI}}^{d}$
gives the final marginal distribution of the demon.

\subsection{Derivation of Eq. (\ref{eq:hie})}

Here, we present the derivation of the hierarchical relation between
distance, namely Eq. (\ref{eq:hie}) in the main text. The distribution
of the interacting bit at the end of the $n$th interaction interval,
are connected to the demon distribution at the start of the $n$th
interaction interval as

\begin{equation}
\bm{p}_{\text{FIN}}^{b,n}=\bm{W}\bm{p}_{\text{INI}}^{d,n},\label{eq:pro}
\end{equation}
where, $\bm{W}=\mathcal{P}^{b}e^{\mathcal{R}\tau}\mathcal{M}$ with

\begin{equation}
\mathcal{P}^{b}=\begin{pmatrix}1 & 1 & 1 & 0 & 0 & 0\\
0 & 0 & 0 & 1 & 1 & 1
\end{pmatrix}
\end{equation}
projecting out the state of the demon. For the bit labeled $0$ and
$1$,

\begin{equation}
p_{\text{FIN},0}^{b,n}=\sum_{i}W_{1i}p_{\text{INI},i}^{d,n},
\end{equation}

\begin{equation}
p_{\text{FIN},1}^{b,n}=\sum_{i}W_{2i}p_{\text{INI},i}^{d,n},
\end{equation}
where $W_{1i}+W_{2i}=1$. In the following, we examine the relationship
between the distance of the bit distribution
\begin{equation}
D(\bm{p}_{\text{FIN}}^{b,n},\bm{p}_{\text{FIN}}^{b,0})=\left|p_{\text{FIN},0}^{b,n}-p_{\text{FIN},0}^{b,0}\right|+\left|p_{\text{FIN},1}^{b,n}-p_{\text{FIN},1}^{b,0}\right|
\end{equation}
and the distance of demon distribution

\begin{equation}
D(\bm{p}_{\text{INI}}^{d,n},\bm{p}_{\text{INI}}^{d,0})=\sum_{i}\left|p_{\text{INI},i}^{d,n}-p_{\text{INI},i}^{d,0}\right|.
\end{equation}
Since

\begin{align}
D(\bm{p}_{\text{FIN}}^{b,n},\bm{p}_{\text{FIN}}^{b,0}) & =\left|\sum_{i}W_{1i}\left(p_{\text{INI},i}^{d,n}-p_{\text{INI},i}^{d,0}\right)\right|+\left|\sum_{i}W_{2i}\left(p_{\text{INI},i}^{d,n}-p_{\text{INI},i}^{d,0}\right)\right|\nonumber \\
 & \le\sum_{i}\left(W_{1i}+W_{2i}\right)\left|p_{\text{INI},i}^{d,n}-p_{\text{INI},i}^{d,0}\right|\nonumber \\
 & =D(\bm{p}_{\text{INI}}^{d,n},\bm{p}_{\text{INI}}^{d,0}),\label{eq:inequ}
\end{align}
which has been presented in the main text as Eq. (\ref{eq:hie}).
The LHS of Eq. (\ref{eq:inequ}) corresponds to the distance between
final bit distribution of the $n$-th interval and initial interval,
and the RHS of Eq. (\ref{eq:inequ}) is the distance between initial
demon distribution of the $n$-th interval and initial interval. Therefore,
the above inequality serves as a hierarchy of the distance function
between the distribution of demon and bit. Such hierarchical relation
follows from Eq. (\ref{eq:pro}), $\bm{p}_{\text{FIN}}^{b,n}=\bm{W}\bm{p}_{\text{INI}}^{d,n}$,
which reveals that the bit distribution $\bm{p}_{\text{FIN}}^{b,n}$
can be identified as the projection of the demon distribution $\bm{p}_{\text{INI}}^{d,n}$
in a lower dimension.

\subsection{Derivation of the speed limit Eq.(\ref{eq:SL})}

With the help of the hierarchical relation between distance, we derive
an explicit speed limit inequality. As mentioned in the main text,
we assume that there is a critical interval number $N_{c}$, satisfying
$\sum_{i=0}|p_{\text{FIN},i}^{b,N_{c}}-p_{\text{FIN},i}^{b,\infty}|\leq d$.
For brevity, we let $p_{\text{INI},i}^{d,n}\equiv p_{i}^{n}$ on the
following derivations.Then
\begin{align}
D(\bm{p}_{\text{INI}}^{d,\infty},\bm{p}_{\text{INI}}^{d,0}) & =\sum_{i}\left|p_{i}^{\infty}-p_{i}^{0}\right|\nonumber \\
 & \leq\sum_{i=0}\left|p_{i}^{0}-p_{i}^{N_{c}}\right|+d\nonumber \\
 & =\sum_{i=0}\left|\sum_{n=0}^{N_{c}}\left(p_{i}^{n+1}-p_{i}^{n}\right)\right|+d\nonumber \\
 & =\sum_{i=0}\left|\sum_{n=0}^{N_{c}}\left(\sum_{j=0}\mathcal{T}_{ij}p_{j}^{n}-p_{i}^{n}\right)\right|+d.
\end{align}
One can define a pseudo average entropy production for the $n$-th
interaction interval as
\begin{align}
\Delta_{n}S\equiv & \sum_{i,j}\mathcal{T}_{ij}p_{j}^{n}\ln\frac{\mathcal{T}_{ij}p_{j}^{n}}{\mathcal{T}_{ji}p_{i}^{n}}\nonumber \\
= & \frac{1}{2}\sum_{i,j}\left(\mathcal{T}_{ij}p_{j}^{n}-\mathcal{T}_{ji}p_{i}^{n}\right)\ln\frac{\mathcal{T}_{ij}p_{j}^{n}}{\mathcal{T}_{ji}p_{i}^{n}}\nonumber \\
\geq & \sum_{i,j}\frac{\left(\mathcal{T}_{ij}p_{j}^{n}-\mathcal{T}_{ji}p_{i}^{n}\right)^{2}}{\mathcal{T}_{ij}p_{j}^{n}+\mathcal{T}_{ji}p_{i}^{n}},
\end{align}
where in the last line the inequality $(a-b)\ln(a/b)\geq2(a-b)^{2}/(a+b)$
has been used. Moreover, to quantify the system's time scale, we introduce
the dynamical activity $A(n)$ and its time average $\langle A\rangle_{N}$
as
\begin{align}
A(n) & \equiv\sum_{i\neq j}\mathcal{T}_{ij}p_{j}^{n}=\sum_{i<j}\left(\mathcal{T}_{ij}p_{j}^{n}+\mathcal{T}_{ji}p_{i}^{n}\right),\\
\langle A\rangle_{N} & \equiv\frac{1}{N}\sum_{n=0}^{N}A(n).
\end{align}
Using the property of the transition matrix $\boldsymbol{\mathcal{T}}$,
$\sum_{j=0}\mathcal{T}_{ji}=1$, the summation term of the total distance
can be rewritten as
\begin{align}
2\Phi-d= & \sum_{i=0}\left|\sum_{n=0}^{N_{c}}\left(\sum_{j=0}\mathcal{T}_{ij}p_{j}^{n}-p_{i}^{n}\right)\right|\nonumber \\
= & \sum_{i=0}\left|\sum_{n=0}^{N_{c}}\left(\sum_{j=0}\mathcal{T}_{ij}p_{j}^{n}-\left(\sum_{j=0}\mathcal{T}_{ji}\right)p_{i}^{n}\right)\right|\nonumber \\
\leq & \sum_{n=0}^{N_{c}}\sum_{i=0}\left|\sum_{j=0}\left(\mathcal{T}_{ij}p_{j}^{n}-\mathcal{T}_{ji}p_{i}^{n}\right)\right|\nonumber \\
\leq & \sum_{n=0}^{N_{c}}\sum_{i=0}\sqrt{\left(\sum_{j=0}\frac{\left(\mathcal{T}_{ij}p_{j}^{n}-\mathcal{T}_{ji}p_{i}^{n}\right)^{2}}{\mathcal{T}_{ij}p_{j}^{n}+\mathcal{T}_{ji}p_{i}^{n}}\right)\left(\sum_{j=0}(\mathcal{T}_{ij}p_{j}^{n}+\mathcal{T}_{ji}p_{i}^{n})\right)}\nonumber \\
\leq & \sum_{n=0}^{N_{c}}\sqrt{\left(\sum_{i,j}\frac{\left(\mathcal{T}_{ij}p_{j}^{n}-\mathcal{T}_{ji}p_{i}^{n}\right)^{2}}{\mathcal{T}_{ij}p_{j}^{n}+\mathcal{T}_{ji}p_{i}^{n}}\right)\left(\sum_{i,j}(\mathcal{T}_{ij}p_{j}^{n}+\mathcal{T}_{ji}p_{i}^{n})\right)}\nonumber \\
\leq & \sum_{n=0}^{N_{c}}\sqrt{\Delta_{n}S\cdot2A(n)}\nonumber \\
\leq & \sqrt{\left(\sum_{n=0}^{N_{c}}\Delta_{n}S\right)\cdot\left(\sum_{n=0}^{N_{c}}2A(n)\right)}=\sqrt{2N_{c}\Sigma_{N_{c}}\langle A\rangle_{N_{c}}},
\end{align}
where $\Sigma_{N_{c}}\equiv\sum_{n=0}^{N_{c}}\Delta_{n}S$ is the
total pseudo average entropy production during these $N_{c}$ intervals.
When $d\to0$, we arrive at $\Phi^{2}\le\frac{1}{2}N_{c}\Sigma_{N_{c}}\langle A\rangle_{N_{c}}.${}
By simply rewriting, a lower bound of the critical interval number
$N_{c}$ is obtained as
\begin{equation}
N_{c}\geq N_{SL}\equiv\frac{\Phi}{\sqrt{\frac{1}{2}\text{\ensuremath{\dot{\Sigma}_{N_{c}}}}\cdot\langle A\rangle_{N_{c}}}},
\end{equation}
which is referred to the Eq. (\ref{eq:SL}) in the main text. Here,
$\dot{\Sigma}_{N_{c}}\equiv N_{c}^{-1}\Sigma$ is the time averaged
of the entropy production per interval (i.e. approximately the entropy
production rate during the relaxation process).

\subsection{Derivation of the tighter speed limit Eq. (\ref{eq:tsl})}

The equality
\begin{equation}
\frac{\left(a-b\right)^{2}}{a+b}=\frac{\left[(a-b)\ln\frac{a}{b}\right]^{2}}{4(a+b)}f\left(\frac{(a-b)\ln\frac{a}{b}}{2(a+b)}\right)^{-2}\label{eq:equality}
\end{equation}
can help us to derive a tighter lower bound for the relaxation interval
number $N_{c}$ than inequality (\ref{eq:SL}), where the concave
function $f(x)$ is the inverse function of $x\tanh(x)$ \cite{vo2022unified,lee2022speed}.
For simplicity, we define the jump frequency $a_{ij}^{n}$, probability
flux $J_{ij}^{n}$ and entropy production rate $\sigma_{ij}^{n}$
associated with the state $i$ and $j$ of the bit as
\begin{align*}
\text{\ensuremath{a_{ij}^{n}\equiv\mathcal{T}_{ij}p_{j}^{n}+\mathcal{T}_{ji}p_{i}^{n},\ }} & J_{ij}^{n}\equiv\mathcal{T}_{ij}p_{j}^{n}-\mathcal{T}_{ji}p_{i}^{n}\\
\sigma_{ij}^{n} & =J_{ij}^{n}\ln\frac{\mathcal{T}_{ij}p_{j}^{n}}{\mathcal{T}_{ji}p_{i}^{n}},
\end{align*}
so that the dynamical activity and the entropy production per interaction
can be rewritten as
\[
A(n)=\sum_{i<j}a_{ij}^{n},\ \Delta_{n}S=\sum_{i<j}\sigma_{ij}^{n}.
\]
According to equation (\ref{eq:equality}), one has
\begin{equation}
\frac{(J_{ij}^{n})^{2}}{a_{ij}^{n}}=\frac{(\sigma_{ij}^{n})^{2}}{4a_{ij}^{n}}f\left(\frac{\sigma_{ij}^{n}}{2a_{ij}^{n}}\right)^{-2},
\end{equation}
then
\begin{align}
2\Phi\leq & \sum_{n=0}^{N_{c}}\sqrt{\left(\sum_{i\neq j}\frac{\left(\mathcal{T}_{ij}p_{j}^{n}-\mathcal{T}_{ji}p_{i}^{n}\right)^{2}}{\mathcal{T}_{ij}p_{j}^{n}+\mathcal{T}_{ji}p_{i}^{n}}\right)\left(\sum_{i\neq j}(\mathcal{T}_{ij}p_{j}^{n}+\mathcal{T}_{ji}p_{i}^{n})\right)}\nonumber \\
= & \sum_{n=0}^{N_{c}}\sqrt{\left(2\sum_{i<j}\frac{\left(J_{ij}^{n}\right)^{2}}{a_{ij}^{n}}\right)\left(2A(n)\right)}\nonumber \\
= & \sum_{n=0}^{N_{c}}\sqrt{\sum_{i<j}\left[\frac{(\sigma_{ij}^{n})^{2}}{2a_{ij}^{n}}f\left(\frac{\sigma_{ij}^{n}}{2a_{ij}^{n}}\right)^{-2}\right]\left(2A(n)\right)}\nonumber \\
\leq & \sum_{n=0}^{N_{c}}\sqrt{\left[\frac{\left(\sum_{i<j}\sigma_{ij}^{n}\right)^{2}}{2\sum_{i<j}a_{ij}^{n}}f\left(\frac{\sum_{i<j}\sigma_{ij}^{n}}{2\sum_{i<j}a_{ij}^{n}}\right)^{-2}\right]\left(2A(n)\right)}\nonumber \\
= & \sum_{n=0}^{N_{c}}\sqrt{\left[\frac{\left(\Delta_{n}S\right)^{2}}{2A(n)}f\left(\frac{\Delta_{n}S}{2A(n)}\right)^{-2}\right]\left(2A(n)\right)}\nonumber \\
\leq & \sqrt{\left[\frac{\left(\sum_{n=0}^{N_{c}}\Delta_{n}S\right)^{2}}{\sum_{n=0}^{N_{c}}2A(n)}f\left(\frac{\sum_{n=0}^{N_{c}}\Delta_{n}S}{\sum_{n=0}^{N_{c}}2A(n)}\right)^{-2}\right]\left(\sum_{n=0}^{N_{c}}2A(n)\right)}\nonumber \\
= & \sqrt{\Sigma^{2}f\left(\frac{\dot{\Sigma}_{N_{c}}}{2\langle A\rangle_{N_{c}}}\right)^{-2}}=\frac{N_{c}\dot{\Sigma}_{N_{c}}}{f\left(\frac{\dot{\Sigma}_{N_{c}}}{2\langle A\rangle_{N_{c}}}\right)},\label{eq:bound2}
\end{align}
thus the new bound is given by

\begin{equation}
N_{c}\geq N_{TSL}=\frac{2\Phi}{\dot{\Sigma}_{N_{c}}}\cdot f\left(\frac{\dot{\Sigma}_{N_{c}}}{2\langle A\rangle_{N_{c}}}\right).\label{eq:TSL}
\end{equation}
It can be observed that
\begin{equation}
N_{TSL}\ge\max\left\{ N_{SL},\frac{\Phi}{\langle A\rangle_{N_{c}}}\right\} ,
\end{equation}
which is always tighter than the first bound $N_{c}\geq N_{SL}$.
Note that in the Eq. (\ref{eq:bound2}), the concavity property of
the function $(x^{2}/y)f(x/y)^{-2}$ for $x,y>0$ and $\tanh(x)<1$
for $x,y>0$ has been used. Various appropriate choices of the concave
function $g(x)$ satisfying the relation

\begin{equation}
\frac{\left(a-b\right)^{2}}{a+b}\le\frac{\left[(a-b)\ln\frac{a}{b}\right]^{2}}{4(a+b)}g\left(\frac{(a-b)\ln\frac{a}{b}}{2(a+b)}\right)^{-2}
\end{equation}
lead to improved bounds than the speed limit inequality, namely Eq.
(\ref{eq:SL}) in the main text. Since $f(x)$ makes the above relation
identity {[}c.f. Eq. (\ref{eq:equality}){]}, Eq. (\ref{eq:TSL})
provides the tightest form of the speed limit inequality.

\section{Analysis of the relaxation modes and timescales}

Here, we provide detailed analysis of the relaxation modes and timescales
based on the spectral decomposition. As shown in the main text, the
transition matrix $\mathcal{\bm{\mathcal{T}}}$ has right eigenvectors
$\bm{R}_{i}$, $\bm{\mathcal{T}}\bm{R}_{i}=\mu_{i}\bm{R}_{i}$ , and
left eigenvectors $\bm{L}_{i}$ as $\bm{L}_{i}^{\text{T}}\bm{\mathcal{T}}=\mu_{i}\bm{L}_{i}^{\text{T}}$
with $\mu_{i}$ the eigenvalues, which are sorted as $1=\mu_{1}>\left|\mu_{2}\right|\ge\left|\mu_{3}\right|\ge...$
. The right eigenvector $\bm{R}_{0}$ with $1=\mu_{0}$ corresponds
to the periodic steady state, so we write $\bm{R}_{1}=\bm{p}_{\text{INI}}^{d,ps}$.
The initial state $\bm{p}_{\text{INI}}^{d,0}$ can be expanded as
$\bm{p}_{\text{INI}}^{d,0}=\bm{p}_{\text{INI}}^{d,ps}+\sum_{i>1}d_{i}\bm{R}_{i}$,
where $d_{i}=\frac{\bm{L}_{i}^{\text{T}}\cdot\bm{p}_{\text{INI}}^{d,0}}{\bm{L}_{i}^{\text{T}}\cdot\bm{R}_{i}}$
{[}c.f. Eq. (\ref{eq:0th}) and (\ref{eq:di}) in the main text{]}.
For an evolution starting at a given initial distribution $\bm{p}_{\text{INI}}^{d}$,
we have that $d_{i}$ is the corresponding overlap coefficient between
the initial probability and the $i$-th relaxation mode, represented
by left eigenvector $\bm{L}_{i}^{\text{T}}$. During the relaxation
process, the initial distribution of the demon of the $n$-th time
interval, $\mathcal{\bm{\mathcal{T}}}^{n}\bm{p}_{\text{INI}}^{d,0}$,
can be written as $\mathcal{\bm{\mathcal{T}}}^{n}\bm{p}_{\text{INI}}^{d,0}=\bm{p}_{\text{INI}}^{d,ps}+\sum_{i>1}d_{i}\mu_{i}^{n}\bm{R}_{i}$
{[}c.f. Eq. (\ref{eq:disn}) in the main text{]}. The distance between
the initial distribution of the demon of the $n$-th time interval
and the periodic steady state can be written as

\begin{equation}
\left\Vert \bm{\mathcal{T}}^{n}\bm{p}_{\text{INI}}^{d,0}-\bm{p}_{\text{INI}}^{d,ps}\right\Vert _{q}=\sum_{i>1}d_{i}\left\Vert \mu_{i}\right\Vert _{q}^{n}\left\Vert R_{i}\right\Vert _{q},
\end{equation}
which reveals that the decay process depends on the relaxation timescales
$\left\{ \mu_{i}\right\} $ and the overlap coefficients $\left\{ d_{i}\right\} $,
especially the dominant elements $\mu_{2}$ and $d_{2}$.

For the three-state model we used, we present the relaxation mode
analysis of the transition matrix $\mathcal{T}_{3\times3}$ , whose
eigenvalues can be solved exactly as

\begin{equation}
\mu_{1}=1,
\end{equation}

\begin{equation}
\mu_{2}=\frac{\sigma\left[\sigma\left(4+\sigma^{\sqrt{3}}+\sigma^{-\sqrt{3}}\right)+\delta\epsilon\left(6-4\sigma-\sigma^{1-\sqrt{3}}-\sigma^{1+\sqrt{3}}\right)\right]}{6},
\end{equation}

\begin{equation}
\mu_{3}=\sigma^{3},
\end{equation}
where $\sigma=e^{-\tau}\in[0,1].$ It has been confirmed by $Mathematica$
that $\mu_{2}$ is always larger than $\mu_{3}$. Hence, one can optimally
design the initial state for faster functionalization, whose overlap
between the relaxation mode corresponds to the timescale $\mu_{2}$
is zero, i.e. $\bm{L}_{2}^{\text{T}}\cdot\bm{\pi}_{\text{INI}}^{d}=0$.
In practice, we can prepare the demon in a thermal reservoir at an
optimal temperature $T_{\text{ini}}=T_{\text{opt}}\neq T$. The initial
demon distribution of each state in equilibrium read

\begin{equation}
\bm{\pi}_{\text{INI}}^{d}(A)=\bm{\pi}_{\text{INI}}^{d}(C)=\frac{e^{-\beta_{\text{opt}}\Delta E}}{1+2e^{-\beta_{\text{opt}}\Delta E}},\qquad\bm{\pi}_{\text{INI}}^{d}(B)=\frac{1}{1+2e^{-\beta_{\text{opt}}\Delta E}},
\end{equation}
where $\beta_{\text{opt}}=1/T_{\text{opt}}$. By solving $L_{21}\bm{\pi}_{\text{INI}}^{d}(A)+L_{22}\bm{\pi}_{\text{INI}}^{d}(B)+L_{23}\bm{\pi}_{\text{INI}}^{d}(C)=0$,
the optimal temperature can be solved as
\begin{equation}
T_{\text{opt}}/\Delta E=\left[k_{b}\ln(-\frac{L_{21}+L_{23}}{L_{22}})\right]^{-1}.
\end{equation}

\bibliographystyle{amsplain}

\begin{thebibliography}{82}
\expandafter\ifx\csname natexlab\endcsname\relax\def\natexlab#1{#1}\fi
\expandafter\ifx\csname bibnamefont\endcsname\relax
  \def\bibnamefont#1{#1}\fi
\expandafter\ifx\csname bibfnamefont\endcsname\relax
  \def\bibfnamefont#1{#1}\fi
\expandafter\ifx\csname citenamefont\endcsname\relax
  \def\citenamefont#1{#1}\fi
\expandafter\ifx\csname url\endcsname\relax
  \def\url#1{\texttt{#1}}\fi
\expandafter\ifx\csname urlprefix\endcsname\relax\def\urlprefix{URL }\fi
\providecommand{\bibinfo}[2]{#2}
\providecommand{\eprint}[2][]{\url{#2}}

\bibitem[{\citenamefont{Maxwell and Pesic}(2001)}]{maxwell2001theory}
\bibinfo{author}{\bibfnamefont{J.~C.} \bibnamefont{Maxwell}} \bibnamefont{and}
  \bibinfo{author}{\bibfnamefont{P.}~\bibnamefont{Pesic}},
  \emph{\bibinfo{title}{Theory of heat}} (\bibinfo{publisher}{Courier
  Corporation}, \bibinfo{year}{2001}).

\bibitem[{\citenamefont{Smoluchowski}(1927)}]{smoluchowski1927experimentell}
\bibinfo{author}{\bibfnamefont{M.}~\bibnamefont{Smoluchowski}},
  \bibinfo{journal}{Pisma Mariana Smoluchowskiego}
  \textbf{\bibinfo{volume}{2}}, \bibinfo{pages}{226} (\bibinfo{year}{1927}).

\bibitem[{\citenamefont{Szilard}(1929)}]{szilard1929entropieverminderung}
\bibinfo{author}{\bibfnamefont{L.}~\bibnamefont{Szilard}},
  \bibinfo{journal}{Zeitschrift f{\"u}r Physik} \textbf{\bibinfo{volume}{53}},
  \bibinfo{pages}{840} (\bibinfo{year}{1929}).

\bibitem[{\citenamefont{Brillouin}(1951)}]{brillouin1951maxwell}
\bibinfo{author}{\bibfnamefont{L.}~\bibnamefont{Brillouin}},
  \bibinfo{journal}{Journal of Applied Physics} \textbf{\bibinfo{volume}{22}},
  \bibinfo{pages}{334} (\bibinfo{year}{1951}).

\bibitem[{\citenamefont{Penrose}(2005)}]{penrose2005foundations}
\bibinfo{author}{\bibfnamefont{O.}~\bibnamefont{Penrose}},
  \emph{\bibinfo{title}{Foundations of statistical mechanics: a deductive
  treatment}} (\bibinfo{publisher}{Courier Corporation}, \bibinfo{year}{2005}).

\bibitem[{\citenamefont{Feynman et~al.}(2011)\citenamefont{Feynman, Leighton,
  and Sands}}]{feynman2011feynman}
\bibinfo{author}{\bibfnamefont{R.~P.} \bibnamefont{Feynman}},
  \bibinfo{author}{\bibfnamefont{R.~B.} \bibnamefont{Leighton}},
  \bibnamefont{and} \bibinfo{author}{\bibfnamefont{M.}~\bibnamefont{Sands}},
  \emph{\bibinfo{title}{The Feynman lectures on physics, Vol. I: The new
  millennium edition: mainly mechanics, radiation, and heat}},
  vol.~\bibinfo{volume}{1} (\bibinfo{publisher}{Basic books},
  \bibinfo{year}{2011}).

\bibitem[{\citenamefont{Landauer}(1961)}]{landauer1961irreversibility}
\bibinfo{author}{\bibfnamefont{R.}~\bibnamefont{Landauer}},
  \bibinfo{journal}{IBM journal of research and development}
  \textbf{\bibinfo{volume}{5}}, \bibinfo{pages}{183} (\bibinfo{year}{1961}).

\bibitem[{\citenamefont{Bennett}(1982)}]{bennett1982thermodynamics}
\bibinfo{author}{\bibfnamefont{C.~H.} \bibnamefont{Bennett}},
  \bibinfo{journal}{International Journal of Theoretical Physics}
  \textbf{\bibinfo{volume}{21}}, \bibinfo{pages}{905} (\bibinfo{year}{1982}).

\bibitem[{\citenamefont{Maruyama et~al.}(2009)\citenamefont{Maruyama, Nori, and
  Vedral}}]{maruyama2009colloquium}
\bibinfo{author}{\bibfnamefont{K.}~\bibnamefont{Maruyama}},
  \bibinfo{author}{\bibfnamefont{F.}~\bibnamefont{Nori}}, \bibnamefont{and}
  \bibinfo{author}{\bibfnamefont{V.}~\bibnamefont{Vedral}},
  \bibinfo{journal}{Reviews of Modern Physics} \textbf{\bibinfo{volume}{81}},
  \bibinfo{pages}{1} (\bibinfo{year}{2009}).

\bibitem[{\citenamefont{Serreli et~al.}(2007)\citenamefont{Serreli, Lee, Kay,
  and Leigh}}]{serreli2007molecular}
\bibinfo{author}{\bibfnamefont{V.}~\bibnamefont{Serreli}},
  \bibinfo{author}{\bibfnamefont{C.-F.} \bibnamefont{Lee}},
  \bibinfo{author}{\bibfnamefont{E.~R.} \bibnamefont{Kay}}, \bibnamefont{and}
  \bibinfo{author}{\bibfnamefont{D.~A.} \bibnamefont{Leigh}},
  \bibinfo{journal}{Nature} \textbf{\bibinfo{volume}{445}},
  \bibinfo{pages}{523} (\bibinfo{year}{2007}).

\bibitem[{\citenamefont{B{\'e}rut et~al.}(2012)\citenamefont{B{\'e}rut,
  Arakelyan, Petrosyan, Ciliberto, Dillenschneider, and
  Lutz}}]{berut2012experimental}
\bibinfo{author}{\bibfnamefont{A.}~\bibnamefont{B{\'e}rut}},
  \bibinfo{author}{\bibfnamefont{A.}~\bibnamefont{Arakelyan}},
  \bibinfo{author}{\bibfnamefont{A.}~\bibnamefont{Petrosyan}},
  \bibinfo{author}{\bibfnamefont{S.}~\bibnamefont{Ciliberto}},
  \bibinfo{author}{\bibfnamefont{R.}~\bibnamefont{Dillenschneider}},
  \bibnamefont{and} \bibinfo{author}{\bibfnamefont{E.}~\bibnamefont{Lutz}},
  \bibinfo{journal}{Nature} \textbf{\bibinfo{volume}{483}},
  \bibinfo{pages}{187} (\bibinfo{year}{2012}).

\bibitem[{\citenamefont{Toyabe et~al.}(2010)\citenamefont{Toyabe, Sagawa, Ueda,
  Muneyuki, and Sano}}]{toyabe2010experimental}
\bibinfo{author}{\bibfnamefont{S.}~\bibnamefont{Toyabe}},
  \bibinfo{author}{\bibfnamefont{T.}~\bibnamefont{Sagawa}},
  \bibinfo{author}{\bibfnamefont{M.}~\bibnamefont{Ueda}},
  \bibinfo{author}{\bibfnamefont{E.}~\bibnamefont{Muneyuki}}, \bibnamefont{and}
  \bibinfo{author}{\bibfnamefont{M.}~\bibnamefont{Sano}},
  \bibinfo{journal}{Nature physics} \textbf{\bibinfo{volume}{6}},
  \bibinfo{pages}{988} (\bibinfo{year}{2010}).

\bibitem[{\citenamefont{Koski et~al.}(2014{\natexlab{a}})\citenamefont{Koski,
  Maisi, Sagawa, and Pekola}}]{koski2014exper}
\bibinfo{author}{\bibfnamefont{J.~V.} \bibnamefont{Koski}},
  \bibinfo{author}{\bibfnamefont{V.~F.} \bibnamefont{Maisi}},
  \bibinfo{author}{\bibfnamefont{T.}~\bibnamefont{Sagawa}}, \bibnamefont{and}
  \bibinfo{author}{\bibfnamefont{J.~P.} \bibnamefont{Pekola}},
  \bibinfo{journal}{Physical review letters} \textbf{\bibinfo{volume}{113}},
  \bibinfo{pages}{030601} (\bibinfo{year}{2014}{\natexlab{a}}).

\bibitem[{\citenamefont{Koski et~al.}(2014{\natexlab{b}})\citenamefont{Koski,
  Maisi, Pekola, and Averin}}]{koski2014experimental}
\bibinfo{author}{\bibfnamefont{J.~V.} \bibnamefont{Koski}},
  \bibinfo{author}{\bibfnamefont{V.~F.} \bibnamefont{Maisi}},
  \bibinfo{author}{\bibfnamefont{J.~P.} \bibnamefont{Pekola}},
  \bibnamefont{and} \bibinfo{author}{\bibfnamefont{D.~V.}
  \bibnamefont{Averin}}, \bibinfo{journal}{Proceedings of the National Academy
  of Sciences} \textbf{\bibinfo{volume}{111}}, \bibinfo{pages}{13786}
  (\bibinfo{year}{2014}{\natexlab{b}}).

\bibitem[{\citenamefont{Koski et~al.}(2015)\citenamefont{Koski, Kutvonen,
  Khaymovich, Ala-Nissila, and Pekola}}]{koski2015chip}
\bibinfo{author}{\bibfnamefont{J.~V.} \bibnamefont{Koski}},
  \bibinfo{author}{\bibfnamefont{A.}~\bibnamefont{Kutvonen}},
  \bibinfo{author}{\bibfnamefont{I.~M.} \bibnamefont{Khaymovich}},
  \bibinfo{author}{\bibfnamefont{T.}~\bibnamefont{Ala-Nissila}},
  \bibnamefont{and} \bibinfo{author}{\bibfnamefont{J.~P.}
  \bibnamefont{Pekola}}, \bibinfo{journal}{Physical review letters}
  \textbf{\bibinfo{volume}{115}}, \bibinfo{pages}{260602}
  (\bibinfo{year}{2015}).

\bibitem[{\citenamefont{Vidrighin et~al.}(2016)\citenamefont{Vidrighin,
  Dahlsten, Barbieri, Kim, Vedral, and Walmsley}}]{vidrighin2016photonic}
\bibinfo{author}{\bibfnamefont{M.~D.} \bibnamefont{Vidrighin}},
  \bibinfo{author}{\bibfnamefont{O.}~\bibnamefont{Dahlsten}},
  \bibinfo{author}{\bibfnamefont{M.}~\bibnamefont{Barbieri}},
  \bibinfo{author}{\bibfnamefont{M.}~\bibnamefont{Kim}},
  \bibinfo{author}{\bibfnamefont{V.}~\bibnamefont{Vedral}}, \bibnamefont{and}
  \bibinfo{author}{\bibfnamefont{I.~A.} \bibnamefont{Walmsley}},
  \bibinfo{journal}{Physical review letters} \textbf{\bibinfo{volume}{116}},
  \bibinfo{pages}{050401} (\bibinfo{year}{2016}).

\bibitem[{\citenamefont{Cottet et~al.}(2017)\citenamefont{Cottet, Jezouin,
  Bretheau, Campagne-Ibarcq, Ficheux, Anders, Auff{\`e}ves, Azouit, Rouchon,
  and Huard}}]{cottet2017observing}
\bibinfo{author}{\bibfnamefont{N.}~\bibnamefont{Cottet}},
  \bibinfo{author}{\bibfnamefont{S.}~\bibnamefont{Jezouin}},
  \bibinfo{author}{\bibfnamefont{L.}~\bibnamefont{Bretheau}},
  \bibinfo{author}{\bibfnamefont{P.}~\bibnamefont{Campagne-Ibarcq}},
  \bibinfo{author}{\bibfnamefont{Q.}~\bibnamefont{Ficheux}},
  \bibinfo{author}{\bibfnamefont{J.}~\bibnamefont{Anders}},
  \bibinfo{author}{\bibfnamefont{A.}~\bibnamefont{Auff{\`e}ves}},
  \bibinfo{author}{\bibfnamefont{R.}~\bibnamefont{Azouit}},
  \bibinfo{author}{\bibfnamefont{P.}~\bibnamefont{Rouchon}}, \bibnamefont{and}
  \bibinfo{author}{\bibfnamefont{B.}~\bibnamefont{Huard}},
  \bibinfo{journal}{Proceedings of the National Academy of Sciences}
  \textbf{\bibinfo{volume}{114}}, \bibinfo{pages}{7561} (\bibinfo{year}{2017}).

\bibitem[{\citenamefont{Kumar et~al.}(2018)\citenamefont{Kumar, Wu, Giraldo,
  and Weiss}}]{kumar2018sorting}
\bibinfo{author}{\bibfnamefont{A.}~\bibnamefont{Kumar}},
  \bibinfo{author}{\bibfnamefont{T.-Y.} \bibnamefont{Wu}},
  \bibinfo{author}{\bibfnamefont{F.}~\bibnamefont{Giraldo}}, \bibnamefont{and}
  \bibinfo{author}{\bibfnamefont{D.~S.} \bibnamefont{Weiss}},
  \bibinfo{journal}{Nature} \textbf{\bibinfo{volume}{561}}, \bibinfo{pages}{83}
  (\bibinfo{year}{2018}).

\bibitem[{\citenamefont{Masuyama et~al.}(2018)\citenamefont{Masuyama, Funo,
  Murashita, Noguchi, Kono, Tabuchi, Yamazaki, Ueda, and
  Nakamura}}]{masuyama2018information}
\bibinfo{author}{\bibfnamefont{Y.}~\bibnamefont{Masuyama}},
  \bibinfo{author}{\bibfnamefont{K.}~\bibnamefont{Funo}},
  \bibinfo{author}{\bibfnamefont{Y.}~\bibnamefont{Murashita}},
  \bibinfo{author}{\bibfnamefont{A.}~\bibnamefont{Noguchi}},
  \bibinfo{author}{\bibfnamefont{S.}~\bibnamefont{Kono}},
  \bibinfo{author}{\bibfnamefont{Y.}~\bibnamefont{Tabuchi}},
  \bibinfo{author}{\bibfnamefont{R.}~\bibnamefont{Yamazaki}},
  \bibinfo{author}{\bibfnamefont{M.}~\bibnamefont{Ueda}}, \bibnamefont{and}
  \bibinfo{author}{\bibfnamefont{Y.}~\bibnamefont{Nakamura}},
  \bibinfo{journal}{Nature communications} \textbf{\bibinfo{volume}{9}},
  \bibinfo{pages}{1} (\bibinfo{year}{2018}).

\bibitem[{\citenamefont{Ribezzi-Crivellari and
  Ritort}(2019)}]{ribezzi2019large}
\bibinfo{author}{\bibfnamefont{M.}~\bibnamefont{Ribezzi-Crivellari}}
  \bibnamefont{and} \bibinfo{author}{\bibfnamefont{F.}~\bibnamefont{Ritort}},
  \bibinfo{journal}{Nature Physics} \textbf{\bibinfo{volume}{15}},
  \bibinfo{pages}{660} (\bibinfo{year}{2019}).

\bibitem[{\citenamefont{Paneru et~al.}(2020)\citenamefont{Paneru, Dutta,
  Sagawa, Tlusty, and Pak}}]{paneru2020efficiency}
\bibinfo{author}{\bibfnamefont{G.}~\bibnamefont{Paneru}},
  \bibinfo{author}{\bibfnamefont{S.}~\bibnamefont{Dutta}},
  \bibinfo{author}{\bibfnamefont{T.}~\bibnamefont{Sagawa}},
  \bibinfo{author}{\bibfnamefont{T.}~\bibnamefont{Tlusty}}, \bibnamefont{and}
  \bibinfo{author}{\bibfnamefont{H.~K.} \bibnamefont{Pak}},
  \bibinfo{journal}{Nature communications} \textbf{\bibinfo{volume}{11}},
  \bibinfo{pages}{1} (\bibinfo{year}{2020}).

\bibitem[{\citenamefont{Zurek}(1989)}]{zurek1989thermodynamic}
\bibinfo{author}{\bibfnamefont{W.~H.} \bibnamefont{Zurek}},
  \bibinfo{journal}{Nature} \textbf{\bibinfo{volume}{341}},
  \bibinfo{pages}{119} (\bibinfo{year}{1989}).

\bibitem[{\citenamefont{Hosoya et~al.}(2011)\citenamefont{Hosoya, Maruyama, and
  Shikano}}]{hosoya2011maxwell}
\bibinfo{author}{\bibfnamefont{A.}~\bibnamefont{Hosoya}},
  \bibinfo{author}{\bibfnamefont{K.}~\bibnamefont{Maruyama}}, \bibnamefont{and}
  \bibinfo{author}{\bibfnamefont{Y.}~\bibnamefont{Shikano}},
  \bibinfo{journal}{Physical Review E} \textbf{\bibinfo{volume}{84}},
  \bibinfo{pages}{061117} (\bibinfo{year}{2011}).

\bibitem[{\citenamefont{Mandal and Jarzynski}(2012)}]{mandal2012work}
\bibinfo{author}{\bibfnamefont{D.}~\bibnamefont{Mandal}} \bibnamefont{and}
  \bibinfo{author}{\bibfnamefont{C.}~\bibnamefont{Jarzynski}},
  \bibinfo{journal}{Proceedings of the National Academy of Sciences}
  \textbf{\bibinfo{volume}{109}}, \bibinfo{pages}{11641}
  (\bibinfo{year}{2012}).

\bibitem[{\citenamefont{Mandal et~al.}(2013)\citenamefont{Mandal, Quan, and
  Jarzynski}}]{mandal2013maxwell}
\bibinfo{author}{\bibfnamefont{D.}~\bibnamefont{Mandal}},
  \bibinfo{author}{\bibfnamefont{H.}~\bibnamefont{Quan}}, \bibnamefont{and}
  \bibinfo{author}{\bibfnamefont{C.}~\bibnamefont{Jarzynski}},
  \bibinfo{journal}{Physical review letters} \textbf{\bibinfo{volume}{111}},
  \bibinfo{pages}{030602} (\bibinfo{year}{2013}).

\bibitem[{\citenamefont{Barato and Seifert}(2014)}]{barato2014unifying}
\bibinfo{author}{\bibfnamefont{A.}~\bibnamefont{Barato}} \bibnamefont{and}
  \bibinfo{author}{\bibfnamefont{U.}~\bibnamefont{Seifert}},
  \bibinfo{journal}{Physical review letters} \textbf{\bibinfo{volume}{112}},
  \bibinfo{pages}{090601} (\bibinfo{year}{2014}).

\bibitem[{\citenamefont{Horowitz and
  Esposito}(2014)}]{horowitz2014thermodynamics}
\bibinfo{author}{\bibfnamefont{J.~M.} \bibnamefont{Horowitz}} \bibnamefont{and}
  \bibinfo{author}{\bibfnamefont{M.}~\bibnamefont{Esposito}},
  \bibinfo{journal}{Physical Review X} \textbf{\bibinfo{volume}{4}},
  \bibinfo{pages}{031015} (\bibinfo{year}{2014}).

\bibitem[{\citenamefont{Strasberg et~al.}(2017)\citenamefont{Strasberg,
  Schaller, Brandes, and Esposito}}]{strasberg2017quantum}
\bibinfo{author}{\bibfnamefont{P.}~\bibnamefont{Strasberg}},
  \bibinfo{author}{\bibfnamefont{G.}~\bibnamefont{Schaller}},
  \bibinfo{author}{\bibfnamefont{T.}~\bibnamefont{Brandes}}, \bibnamefont{and}
  \bibinfo{author}{\bibfnamefont{M.}~\bibnamefont{Esposito}},
  \bibinfo{journal}{Physical Review X} \textbf{\bibinfo{volume}{7}},
  \bibinfo{pages}{021003} (\bibinfo{year}{2017}).

\bibitem[{\citenamefont{Joseph and Kiran}(2021)}]{joseph2021efficiency}
\bibinfo{author}{\bibfnamefont{T.}~\bibnamefont{Joseph}} \bibnamefont{and}
  \bibinfo{author}{\bibfnamefont{V.}~\bibnamefont{Kiran}},
  \bibinfo{journal}{Physical Review E} \textbf{\bibinfo{volume}{103}},
  \bibinfo{pages}{022131} (\bibinfo{year}{2021}).

\bibitem[{\citenamefont{Lu and Jarzynski}(2019)}]{lu2019programmable}
\bibinfo{author}{\bibfnamefont{Z.}~\bibnamefont{Lu}} \bibnamefont{and}
  \bibinfo{author}{\bibfnamefont{C.}~\bibnamefont{Jarzynski}},
  \bibinfo{journal}{Entropy} \textbf{\bibinfo{volume}{21}}, \bibinfo{pages}{65}
  (\bibinfo{year}{2019}).

\bibitem[{\citenamefont{Quan et~al.}(2006)\citenamefont{Quan, Wang, Liu, Sun,
  and Nori}}]{quan2006maxwell}
\bibinfo{author}{\bibfnamefont{H.}~\bibnamefont{Quan}},
  \bibinfo{author}{\bibfnamefont{Y.}~\bibnamefont{Wang}},
  \bibinfo{author}{\bibfnamefont{Y.-x.} \bibnamefont{Liu}},
  \bibinfo{author}{\bibfnamefont{C.}~\bibnamefont{Sun}}, \bibnamefont{and}
  \bibinfo{author}{\bibfnamefont{F.}~\bibnamefont{Nori}},
  \bibinfo{journal}{Physical review letters} \textbf{\bibinfo{volume}{97}},
  \bibinfo{pages}{180402} (\bibinfo{year}{2006}).

\bibitem[{\citenamefont{Abreu and Seifert}(2011)}]{abreu2011extracting}
\bibinfo{author}{\bibfnamefont{D.}~\bibnamefont{Abreu}} \bibnamefont{and}
  \bibinfo{author}{\bibfnamefont{U.}~\bibnamefont{Seifert}},
  \bibinfo{journal}{EPL (Europhysics Letters)} \textbf{\bibinfo{volume}{94}},
  \bibinfo{pages}{10001} (\bibinfo{year}{2011}).

\bibitem[{\citenamefont{Aristotle and
  Aristotle}(1933)}]{aristotle1933metaphysics}
\bibinfo{author}{\bibfnamefont{A.}~\bibnamefont{Aristotle}} \bibnamefont{and}
  \bibinfo{author}{\bibnamefont{Aristotle}},
  \emph{\bibinfo{title}{Metaphysics}}, vol.~\bibinfo{volume}{2}
  (\bibinfo{publisher}{Harvard University Press Cambridge, MA},
  \bibinfo{year}{1933}).

\bibitem[{\citenamefont{Mpemba and Osborne}(1969)}]{mpemba1969cool}
\bibinfo{author}{\bibfnamefont{E.~B.} \bibnamefont{Mpemba}} \bibnamefont{and}
  \bibinfo{author}{\bibfnamefont{D.~G.} \bibnamefont{Osborne}},
  \bibinfo{journal}{Physics Education} \textbf{\bibinfo{volume}{4}},
  \bibinfo{pages}{172} (\bibinfo{year}{1969}).

\bibitem[{\citenamefont{Lu and Raz}(2017)}]{lu2017nonequilibrium}
\bibinfo{author}{\bibfnamefont{Z.}~\bibnamefont{Lu}} \bibnamefont{and}
  \bibinfo{author}{\bibfnamefont{O.}~\bibnamefont{Raz}},
  \bibinfo{journal}{Proceedings of the National Academy of Sciences}
  \textbf{\bibinfo{volume}{114}}, \bibinfo{pages}{5083} (\bibinfo{year}{2017}).

\bibitem[{\citenamefont{Klich et~al.}(2019)\citenamefont{Klich, Raz,
  Hirschberg, and Vucelja}}]{klich2019mpemba}
\bibinfo{author}{\bibfnamefont{I.}~\bibnamefont{Klich}},
  \bibinfo{author}{\bibfnamefont{O.}~\bibnamefont{Raz}},
  \bibinfo{author}{\bibfnamefont{O.}~\bibnamefont{Hirschberg}},
  \bibnamefont{and} \bibinfo{author}{\bibfnamefont{M.}~\bibnamefont{Vucelja}},
  \bibinfo{journal}{Physical Review X} \textbf{\bibinfo{volume}{9}},
  \bibinfo{pages}{021060} (\bibinfo{year}{2019}).

\bibitem[{\citenamefont{Gal and Raz}(2020)}]{gal2020precooling}
\bibinfo{author}{\bibfnamefont{A.}~\bibnamefont{Gal}} \bibnamefont{and}
  \bibinfo{author}{\bibfnamefont{O.}~\bibnamefont{Raz}},
  \bibinfo{journal}{Physical review letters} \textbf{\bibinfo{volume}{124}},
  \bibinfo{pages}{060602} (\bibinfo{year}{2020}).

\bibitem[{\citenamefont{Carollo et~al.}(2021)\citenamefont{Carollo, Lasanta,
  and Lesanovsky}}]{carollo2021exponentially}
\bibinfo{author}{\bibfnamefont{F.}~\bibnamefont{Carollo}},
  \bibinfo{author}{\bibfnamefont{A.}~\bibnamefont{Lasanta}}, \bibnamefont{and}
  \bibinfo{author}{\bibfnamefont{I.}~\bibnamefont{Lesanovsky}},
  \bibinfo{journal}{Physical Review Letters} \textbf{\bibinfo{volume}{127}},
  \bibinfo{pages}{060401} (\bibinfo{year}{2021}).

\bibitem[{\citenamefont{Kumar and Bechhoefer}(2020)}]{kumar2020exponentially}
\bibinfo{author}{\bibfnamefont{A.}~\bibnamefont{Kumar}} \bibnamefont{and}
  \bibinfo{author}{\bibfnamefont{J.}~\bibnamefont{Bechhoefer}},
  \bibinfo{journal}{Nature} \textbf{\bibinfo{volume}{584}}, \bibinfo{pages}{64}
  (\bibinfo{year}{2020}).

\bibitem[{\citenamefont{Lasanta et~al.}(2017)\citenamefont{Lasanta, Reyes,
  Prados, and Santos}}]{lasanta2017hotter}
\bibinfo{author}{\bibfnamefont{A.}~\bibnamefont{Lasanta}},
  \bibinfo{author}{\bibfnamefont{F.~V.} \bibnamefont{Reyes}},
  \bibinfo{author}{\bibfnamefont{A.}~\bibnamefont{Prados}}, \bibnamefont{and}
  \bibinfo{author}{\bibfnamefont{A.}~\bibnamefont{Santos}},
  \bibinfo{journal}{Physical review letters} \textbf{\bibinfo{volume}{119}},
  \bibinfo{pages}{148001} (\bibinfo{year}{2017}).

\bibitem[{\citenamefont{Baity-Jesi et~al.}(2019)\citenamefont{Baity-Jesi,
  Calore, Cruz, Fernandez, Gil-Narvi{\'o}n, Gordillo-Guerrero, I{\~n}iguez,
  Lasanta, Maiorano, Marinari et~al.}}]{baity2019mpemba}
\bibinfo{author}{\bibfnamefont{M.}~\bibnamefont{Baity-Jesi}},
  \bibinfo{author}{\bibfnamefont{E.}~\bibnamefont{Calore}},
  \bibinfo{author}{\bibfnamefont{A.}~\bibnamefont{Cruz}},
  \bibinfo{author}{\bibfnamefont{L.~A.} \bibnamefont{Fernandez}},
  \bibinfo{author}{\bibfnamefont{J.~M.} \bibnamefont{Gil-Narvi{\'o}n}},
  \bibinfo{author}{\bibfnamefont{A.}~\bibnamefont{Gordillo-Guerrero}},
  \bibinfo{author}{\bibfnamefont{D.}~\bibnamefont{I{\~n}iguez}},
  \bibinfo{author}{\bibfnamefont{A.}~\bibnamefont{Lasanta}},
  \bibinfo{author}{\bibfnamefont{A.}~\bibnamefont{Maiorano}},
  \bibinfo{author}{\bibfnamefont{E.}~\bibnamefont{Marinari}},
  \bibnamefont{et~al.}, \bibinfo{journal}{Proceedings of the National Academy
  of Sciences} \textbf{\bibinfo{volume}{116}}, \bibinfo{pages}{15350}
  (\bibinfo{year}{2019}).

\bibitem[{\citenamefont{Gij{\'o}n et~al.}(2019)\citenamefont{Gij{\'o}n,
  Lasanta, and Hern{\'a}ndez}}]{gijon2019paths}
\bibinfo{author}{\bibfnamefont{A.}~\bibnamefont{Gij{\'o}n}},
  \bibinfo{author}{\bibfnamefont{A.}~\bibnamefont{Lasanta}}, \bibnamefont{and}
  \bibinfo{author}{\bibfnamefont{E.}~\bibnamefont{Hern{\'a}ndez}},
  \bibinfo{journal}{Physical Review E} \textbf{\bibinfo{volume}{100}},
  \bibinfo{pages}{032103} (\bibinfo{year}{2019}).

\bibitem[{\citenamefont{Torrente et~al.}(2019)\citenamefont{Torrente,
  L{\'o}pez-Casta{\~n}o, Lasanta, Reyes, Prados, and
  Santos}}]{torrente2019large}
\bibinfo{author}{\bibfnamefont{A.}~\bibnamefont{Torrente}},
  \bibinfo{author}{\bibfnamefont{M.~A.} \bibnamefont{L{\'o}pez-Casta{\~n}o}},
  \bibinfo{author}{\bibfnamefont{A.}~\bibnamefont{Lasanta}},
  \bibinfo{author}{\bibfnamefont{F.~V.} \bibnamefont{Reyes}},
  \bibinfo{author}{\bibfnamefont{A.}~\bibnamefont{Prados}}, \bibnamefont{and}
  \bibinfo{author}{\bibfnamefont{A.}~\bibnamefont{Santos}},
  \bibinfo{journal}{Physical Review E} \textbf{\bibinfo{volume}{99}},
  \bibinfo{pages}{060901} (\bibinfo{year}{2019}).

\bibitem[{\citenamefont{Ch{\'e}trite et~al.}(2021)\citenamefont{Ch{\'e}trite,
  Kumar, and Bechhoefer}}]{chetrite2021metastable}
\bibinfo{author}{\bibfnamefont{R.}~\bibnamefont{Ch{\'e}trite}},
  \bibinfo{author}{\bibfnamefont{A.}~\bibnamefont{Kumar}}, \bibnamefont{and}
  \bibinfo{author}{\bibfnamefont{J.}~\bibnamefont{Bechhoefer}},
  \bibinfo{journal}{Frontiers in Physics} \textbf{\bibinfo{volume}{9}},
  \bibinfo{pages}{141} (\bibinfo{year}{2021}).

\bibitem[{\citenamefont{Vadakkayil and Das}(2021)}]{vadakkayil2021should}
\bibinfo{author}{\bibfnamefont{N.}~\bibnamefont{Vadakkayil}} \bibnamefont{and}
  \bibinfo{author}{\bibfnamefont{S.~K.} \bibnamefont{Das}},
  \bibinfo{journal}{Physical Chemistry Chemical Physics}
  \textbf{\bibinfo{volume}{23}}, \bibinfo{pages}{11186} (\bibinfo{year}{2021}).

\bibitem[{\citenamefont{Yang and Hou}(2020)}]{yang2020non}
\bibinfo{author}{\bibfnamefont{Z.-Y.} \bibnamefont{Yang}} \bibnamefont{and}
  \bibinfo{author}{\bibfnamefont{J.-X.} \bibnamefont{Hou}},
  \bibinfo{journal}{Physical Review E} \textbf{\bibinfo{volume}{101}},
  \bibinfo{pages}{052106} (\bibinfo{year}{2020}).

\bibitem[{\citenamefont{Busiello et~al.}(2021)\citenamefont{Busiello, Gupta,
  and Maritan}}]{busiello2021inducing}
\bibinfo{author}{\bibfnamefont{D.~M.} \bibnamefont{Busiello}},
  \bibinfo{author}{\bibfnamefont{D.}~\bibnamefont{Gupta}}, \bibnamefont{and}
  \bibinfo{author}{\bibfnamefont{A.}~\bibnamefont{Maritan}},
  \bibinfo{journal}{New Journal of Physics} \textbf{\bibinfo{volume}{23}},
  \bibinfo{pages}{103012} (\bibinfo{year}{2021}).

\bibitem[{\citenamefont{Schwarzendahl and
  L{\"o}wen}(2021)}]{schwarzendahl2021anomalous}
\bibinfo{author}{\bibfnamefont{F.~J.} \bibnamefont{Schwarzendahl}}
  \bibnamefont{and}
  \bibinfo{author}{\bibfnamefont{H.}~\bibnamefont{L{\"o}wen}},
  \bibinfo{journal}{arXiv preprint arXiv:2111.06109}  (\bibinfo{year}{2021}).

\bibitem[{\citenamefont{Santos and Prados}(2020)}]{santos2020mpemba}
\bibinfo{author}{\bibfnamefont{A.}~\bibnamefont{Santos}} \bibnamefont{and}
  \bibinfo{author}{\bibfnamefont{A.}~\bibnamefont{Prados}},
  \bibinfo{journal}{Physics of Fluids} \textbf{\bibinfo{volume}{32}},
  \bibinfo{pages}{072010} (\bibinfo{year}{2020}).

\bibitem[{\citenamefont{Biswas et~al.}(2022)\citenamefont{Biswas, Prasad, and
  Rajesh}}]{biswas2022mpemba}
\bibinfo{author}{\bibfnamefont{A.}~\bibnamefont{Biswas}},
  \bibinfo{author}{\bibfnamefont{V.}~\bibnamefont{Prasad}}, \bibnamefont{and}
  \bibinfo{author}{\bibfnamefont{R.}~\bibnamefont{Rajesh}},
  \bibinfo{journal}{Journal of Statistical Physics}
  \textbf{\bibinfo{volume}{186}}, \bibinfo{pages}{1} (\bibinfo{year}{2022}).

\bibitem[{\citenamefont{Shiraishi et~al.}(2018)\citenamefont{Shiraishi, Funo,
  and Saito}}]{shiraishi2018speed}
\bibinfo{author}{\bibfnamefont{N.}~\bibnamefont{Shiraishi}},
  \bibinfo{author}{\bibfnamefont{K.}~\bibnamefont{Funo}}, \bibnamefont{and}
  \bibinfo{author}{\bibfnamefont{K.}~\bibnamefont{Saito}},
  \bibinfo{journal}{Physical review letters} \textbf{\bibinfo{volume}{121}},
  \bibinfo{pages}{070601} (\bibinfo{year}{2018}).

\bibitem[{\citenamefont{Lecomte et~al.}(2007)\citenamefont{Lecomte,
  Appert-Rolland, and Van~Wijland}}]{lecomte2007thermodynamic}
\bibinfo{author}{\bibfnamefont{V.}~\bibnamefont{Lecomte}},
  \bibinfo{author}{\bibfnamefont{C.}~\bibnamefont{Appert-Rolland}},
  \bibnamefont{and}
  \bibinfo{author}{\bibfnamefont{F.}~\bibnamefont{Van~Wijland}},
  \bibinfo{journal}{Journal of statistical physics}
  \textbf{\bibinfo{volume}{127}}, \bibinfo{pages}{51} (\bibinfo{year}{2007}).

\bibitem[{\citenamefont{Garrahan et~al.}(2007)\citenamefont{Garrahan, Jack,
  Lecomte, Pitard, van Duijvendijk, and van Wijland}}]{garrahan2007dynamical}
\bibinfo{author}{\bibfnamefont{J.~P.} \bibnamefont{Garrahan}},
  \bibinfo{author}{\bibfnamefont{R.~L.} \bibnamefont{Jack}},
  \bibinfo{author}{\bibfnamefont{V.}~\bibnamefont{Lecomte}},
  \bibinfo{author}{\bibfnamefont{E.}~\bibnamefont{Pitard}},
  \bibinfo{author}{\bibfnamefont{K.}~\bibnamefont{van Duijvendijk}},
  \bibnamefont{and} \bibinfo{author}{\bibfnamefont{F.}~\bibnamefont{van
  Wijland}}, \bibinfo{journal}{Physical review letters}
  \textbf{\bibinfo{volume}{98}}, \bibinfo{pages}{195702}
  (\bibinfo{year}{2007}).

\bibitem[{\citenamefont{Baiesi et~al.}(2009{\natexlab{a}})\citenamefont{Baiesi,
  Maes, and Wynants}}]{baiesi2009fluctuations}
\bibinfo{author}{\bibfnamefont{M.}~\bibnamefont{Baiesi}},
  \bibinfo{author}{\bibfnamefont{C.}~\bibnamefont{Maes}}, \bibnamefont{and}
  \bibinfo{author}{\bibfnamefont{B.}~\bibnamefont{Wynants}},
  \bibinfo{journal}{Physical review letters} \textbf{\bibinfo{volume}{103}},
  \bibinfo{pages}{010602} (\bibinfo{year}{2009}{\natexlab{a}}).

\bibitem[{\citenamefont{Baiesi et~al.}(2009{\natexlab{b}})\citenamefont{Baiesi,
  Maes, and Wynants}}]{baiesi2009nonequilibrium}
\bibinfo{author}{\bibfnamefont{M.}~\bibnamefont{Baiesi}},
  \bibinfo{author}{\bibfnamefont{C.}~\bibnamefont{Maes}}, \bibnamefont{and}
  \bibinfo{author}{\bibfnamefont{B.}~\bibnamefont{Wynants}},
  \bibinfo{journal}{Journal of statistical physics}
  \textbf{\bibinfo{volume}{137}}, \bibinfo{pages}{1094}
  (\bibinfo{year}{2009}{\natexlab{b}}).

\bibitem[{\citenamefont{Maes}(2017)}]{maes2017non}
\bibinfo{author}{\bibfnamefont{C.}~\bibnamefont{Maes}},
  \emph{\bibinfo{title}{Non-dissipative effects in nonequilibrium systems}}
  (\bibinfo{publisher}{Springer}, \bibinfo{year}{2017}).

\bibitem[{\citenamefont{Di~Terlizzi and Baiesi}(2018)}]{di2018kinetic}
\bibinfo{author}{\bibfnamefont{I.}~\bibnamefont{Di~Terlizzi}} \bibnamefont{and}
  \bibinfo{author}{\bibfnamefont{M.}~\bibnamefont{Baiesi}},
  \bibinfo{journal}{Journal of Physics A: Mathematical and Theoretical}
  \textbf{\bibinfo{volume}{52}}, \bibinfo{pages}{02LT03}
  (\bibinfo{year}{2018}).

\bibitem[{\citenamefont{Mandelstam and Tamm}(1991)}]{mandelstam1991uncertainty}
\bibinfo{author}{\bibfnamefont{L.}~\bibnamefont{Mandelstam}} \bibnamefont{and}
  \bibinfo{author}{\bibfnamefont{I.}~\bibnamefont{Tamm}}, in
  \emph{\bibinfo{booktitle}{Selected papers}} (\bibinfo{publisher}{Springer},
  \bibinfo{year}{1991}), pp. \bibinfo{pages}{115--123}.

\bibitem[{\citenamefont{Fleming}(1973)}]{fleming1973unitarity}
\bibinfo{author}{\bibfnamefont{G.~N.} \bibnamefont{Fleming}},
  \bibinfo{journal}{Il Nuovo Cimento A (1965-1970)}
  \textbf{\bibinfo{volume}{16}}, \bibinfo{pages}{232} (\bibinfo{year}{1973}).

\bibitem[{\citenamefont{Anandan and Aharonov}(1990)}]{anandan1990geometry}
\bibinfo{author}{\bibfnamefont{J.}~\bibnamefont{Anandan}} \bibnamefont{and}
  \bibinfo{author}{\bibfnamefont{Y.}~\bibnamefont{Aharonov}},
  \bibinfo{journal}{Physical review letters} \textbf{\bibinfo{volume}{65}},
  \bibinfo{pages}{1697} (\bibinfo{year}{1990}).

\bibitem[{\citenamefont{Margolus and Levitin}(1998)}]{margolus1998maximum}
\bibinfo{author}{\bibfnamefont{N.}~\bibnamefont{Margolus}} \bibnamefont{and}
  \bibinfo{author}{\bibfnamefont{L.~B.} \bibnamefont{Levitin}},
  \bibinfo{journal}{Physica D: Nonlinear Phenomena}
  \textbf{\bibinfo{volume}{120}}, \bibinfo{pages}{188} (\bibinfo{year}{1998}).

\bibitem[{\citenamefont{Pfeifer}(1993)}]{pfeifer1993fast}
\bibinfo{author}{\bibfnamefont{P.}~\bibnamefont{Pfeifer}},
  \bibinfo{journal}{Physical review letters} \textbf{\bibinfo{volume}{70}},
  \bibinfo{pages}{3365} (\bibinfo{year}{1993}).

\bibitem[{\citenamefont{Taddei et~al.}(2013)\citenamefont{Taddei, Escher,
  Davidovich, and de~Matos~Filho}}]{taddei2013quantum}
\bibinfo{author}{\bibfnamefont{M.~M.} \bibnamefont{Taddei}},
  \bibinfo{author}{\bibfnamefont{B.~M.} \bibnamefont{Escher}},
  \bibinfo{author}{\bibfnamefont{L.}~\bibnamefont{Davidovich}},
  \bibnamefont{and} \bibinfo{author}{\bibfnamefont{R.~L.}
  \bibnamefont{de~Matos~Filho}}, \bibinfo{journal}{Physical review letters}
  \textbf{\bibinfo{volume}{110}}, \bibinfo{pages}{050402}
  (\bibinfo{year}{2013}).

\bibitem[{\citenamefont{del Campo et~al.}(2013)\citenamefont{del Campo,
  Egusquiza, Plenio, and Huelga}}]{del2013quantum}
\bibinfo{author}{\bibfnamefont{A.}~\bibnamefont{del Campo}},
  \bibinfo{author}{\bibfnamefont{I.~L.} \bibnamefont{Egusquiza}},
  \bibinfo{author}{\bibfnamefont{M.~B.} \bibnamefont{Plenio}},
  \bibnamefont{and} \bibinfo{author}{\bibfnamefont{S.~F.}
  \bibnamefont{Huelga}}, \bibinfo{journal}{Physical review letters}
  \textbf{\bibinfo{volume}{110}}, \bibinfo{pages}{050403}
  (\bibinfo{year}{2013}).

\bibitem[{\citenamefont{Deffner and Lutz}(2013)}]{deffner2013quantum}
\bibinfo{author}{\bibfnamefont{S.}~\bibnamefont{Deffner}} \bibnamefont{and}
  \bibinfo{author}{\bibfnamefont{E.}~\bibnamefont{Lutz}},
  \bibinfo{journal}{Physical review letters} \textbf{\bibinfo{volume}{111}},
  \bibinfo{pages}{010402} (\bibinfo{year}{2013}).

\bibitem[{\citenamefont{Pires et~al.}(2016)\citenamefont{Pires, Cianciaruso,
  C{\'e}leri, Adesso, and Soares-Pinto}}]{pires2016generalized}
\bibinfo{author}{\bibfnamefont{D.~P.} \bibnamefont{Pires}},
  \bibinfo{author}{\bibfnamefont{M.}~\bibnamefont{Cianciaruso}},
  \bibinfo{author}{\bibfnamefont{L.~C.} \bibnamefont{C{\'e}leri}},
  \bibinfo{author}{\bibfnamefont{G.}~\bibnamefont{Adesso}}, \bibnamefont{and}
  \bibinfo{author}{\bibfnamefont{D.~O.} \bibnamefont{Soares-Pinto}},
  \bibinfo{journal}{Physical Review X} \textbf{\bibinfo{volume}{6}},
  \bibinfo{pages}{021031} (\bibinfo{year}{2016}).

\bibitem[{\citenamefont{Funo et~al.}(2017)\citenamefont{Funo, Zhang, Chatou,
  Kim, Ueda, and Del~Campo}}]{funo2017universal}
\bibinfo{author}{\bibfnamefont{K.}~\bibnamefont{Funo}},
  \bibinfo{author}{\bibfnamefont{J.-N.} \bibnamefont{Zhang}},
  \bibinfo{author}{\bibfnamefont{C.}~\bibnamefont{Chatou}},
  \bibinfo{author}{\bibfnamefont{K.}~\bibnamefont{Kim}},
  \bibinfo{author}{\bibfnamefont{M.}~\bibnamefont{Ueda}}, \bibnamefont{and}
  \bibinfo{author}{\bibfnamefont{A.}~\bibnamefont{Del~Campo}},
  \bibinfo{journal}{Physical Review Letters} \textbf{\bibinfo{volume}{118}},
  \bibinfo{pages}{100602} (\bibinfo{year}{2017}).

\bibitem[{\citenamefont{Deffner}(2017)}]{deffner2017geometric}
\bibinfo{author}{\bibfnamefont{S.}~\bibnamefont{Deffner}},
  \bibinfo{journal}{New Journal of Physics} \textbf{\bibinfo{volume}{19}},
  \bibinfo{pages}{103018} (\bibinfo{year}{2017}).

\bibitem[{\citenamefont{Shanahan et~al.}(2018)\citenamefont{Shanahan, Chenu,
  Margolus, and Del~Campo}}]{shanahan2018quantum}
\bibinfo{author}{\bibfnamefont{B.}~\bibnamefont{Shanahan}},
  \bibinfo{author}{\bibfnamefont{A.}~\bibnamefont{Chenu}},
  \bibinfo{author}{\bibfnamefont{N.}~\bibnamefont{Margolus}}, \bibnamefont{and}
  \bibinfo{author}{\bibfnamefont{A.}~\bibnamefont{Del~Campo}},
  \bibinfo{journal}{Physical review letters} \textbf{\bibinfo{volume}{120}},
  \bibinfo{pages}{070401} (\bibinfo{year}{2018}).

\bibitem[{\citenamefont{Okuyama and Ohzeki}(2018)}]{okuyama2018quantum}
\bibinfo{author}{\bibfnamefont{M.}~\bibnamefont{Okuyama}} \bibnamefont{and}
  \bibinfo{author}{\bibfnamefont{M.}~\bibnamefont{Ohzeki}},
  \bibinfo{journal}{Physical review letters} \textbf{\bibinfo{volume}{120}},
  \bibinfo{pages}{070402} (\bibinfo{year}{2018}).

\bibitem[{\citenamefont{Ito}(2018)}]{ito2018stochastic}
\bibinfo{author}{\bibfnamefont{S.}~\bibnamefont{Ito}},
  \bibinfo{journal}{Physical review letters} \textbf{\bibinfo{volume}{121}},
  \bibinfo{pages}{030605} (\bibinfo{year}{2018}).

\bibitem[{\citenamefont{Shiraishi and Saito}(2019)}]{shiraishi2019information}
\bibinfo{author}{\bibfnamefont{N.}~\bibnamefont{Shiraishi}} \bibnamefont{and}
  \bibinfo{author}{\bibfnamefont{K.}~\bibnamefont{Saito}},
  \bibinfo{journal}{Physical review letters} \textbf{\bibinfo{volume}{123}},
  \bibinfo{pages}{110603} (\bibinfo{year}{2019}).

\bibitem[{\citenamefont{Nicholson et~al.}(2020)\citenamefont{Nicholson,
  Garcia-Pintos, del Campo, and Green}}]{nicholson2020time}
\bibinfo{author}{\bibfnamefont{S.~B.} \bibnamefont{Nicholson}},
  \bibinfo{author}{\bibfnamefont{L.~P.} \bibnamefont{Garcia-Pintos}},
  \bibinfo{author}{\bibfnamefont{A.}~\bibnamefont{del Campo}},
  \bibnamefont{and} \bibinfo{author}{\bibfnamefont{J.~R.} \bibnamefont{Green}},
  \bibinfo{journal}{Nature Physics} \textbf{\bibinfo{volume}{16}},
  \bibinfo{pages}{1211} (\bibinfo{year}{2020}).

\bibitem[{\citenamefont{Ito and Dechant}(2020)}]{ito2020stochastic}
\bibinfo{author}{\bibfnamefont{S.}~\bibnamefont{Ito}} \bibnamefont{and}
  \bibinfo{author}{\bibfnamefont{A.}~\bibnamefont{Dechant}},
  \bibinfo{journal}{Physical Review X} \textbf{\bibinfo{volume}{10}},
  \bibinfo{pages}{021056} (\bibinfo{year}{2020}).

\bibitem[{\citenamefont{Van~Vu et~al.}(2020)\citenamefont{Van~Vu, Hasegawa
  et~al.}}]{van2020unified}
\bibinfo{author}{\bibfnamefont{T.}~\bibnamefont{Van~Vu}},
  \bibinfo{author}{\bibfnamefont{Y.}~\bibnamefont{Hasegawa}},
  \bibnamefont{et~al.}, \bibinfo{journal}{Physical Review E}
  \textbf{\bibinfo{volume}{102}}, \bibinfo{pages}{062132}
  (\bibinfo{year}{2020}).

\bibitem[{\citenamefont{Gupta and Busiello}(2020)}]{gupta2020tighter}
\bibinfo{author}{\bibfnamefont{D.}~\bibnamefont{Gupta}} \bibnamefont{and}
  \bibinfo{author}{\bibfnamefont{D.~M.} \bibnamefont{Busiello}},
  \bibinfo{journal}{Physical Review E} \textbf{\bibinfo{volume}{102}},
  \bibinfo{pages}{062121} (\bibinfo{year}{2020}).

\bibitem[{\citenamefont{Yoshimura and Ito}(2021)}]{yoshimura2021thermodynamic}
\bibinfo{author}{\bibfnamefont{K.}~\bibnamefont{Yoshimura}} \bibnamefont{and}
  \bibinfo{author}{\bibfnamefont{S.}~\bibnamefont{Ito}},
  \bibinfo{journal}{Physical review letters} \textbf{\bibinfo{volume}{127}},
  \bibinfo{pages}{160601} (\bibinfo{year}{2021}).

\bibitem[{\citenamefont{Vo et~al.}(2022)\citenamefont{Vo, Van~Vu, and
  Hasegawa}}]{vo2022unified}
\bibinfo{author}{\bibfnamefont{V.~T.} \bibnamefont{Vo}},
  \bibinfo{author}{\bibfnamefont{T.}~\bibnamefont{Van~Vu}}, \bibnamefont{and}
  \bibinfo{author}{\bibfnamefont{Y.}~\bibnamefont{Hasegawa}},
  \bibinfo{journal}{arXiv preprint arXiv:2203.11501}  (\bibinfo{year}{2022}).

\bibitem[{\citenamefont{Lee et~al.}(2022)\citenamefont{Lee, Lee, Kwon, and
  Park}}]{lee2022speed}
\bibinfo{author}{\bibfnamefont{J.~S.} \bibnamefont{Lee}},
  \bibinfo{author}{\bibfnamefont{S.}~\bibnamefont{Lee}},
  \bibinfo{author}{\bibfnamefont{H.}~\bibnamefont{Kwon}}, \bibnamefont{and}
  \bibinfo{author}{\bibfnamefont{H.}~\bibnamefont{Park}},
  \bibinfo{journal}{arXiv preprint arXiv:2204.07388}  (\bibinfo{year}{2022}).

\bibitem[{\citenamefont{Hatano and Sasa}(2001)}]{hatano2001steady}
\bibinfo{author}{\bibfnamefont{T.}~\bibnamefont{Hatano}} \bibnamefont{and}
  \bibinfo{author}{\bibfnamefont{S.-i.} \bibnamefont{Sasa}},
  \bibinfo{journal}{Physical review letters} \textbf{\bibinfo{volume}{86}},
  \bibinfo{pages}{3463} (\bibinfo{year}{2001}).

\bibitem[{\citenamefont{Liu et~al.}(2020)\citenamefont{Liu, Gong, and
  Ueda}}]{liu2020thermodynamic}
\bibinfo{author}{\bibfnamefont{K.}~\bibnamefont{Liu}},
  \bibinfo{author}{\bibfnamefont{Z.}~\bibnamefont{Gong}}, \bibnamefont{and}
  \bibinfo{author}{\bibfnamefont{M.}~\bibnamefont{Ueda}},
  \bibinfo{journal}{Physical Review Letters} \textbf{\bibinfo{volume}{125}},
  \bibinfo{pages}{140602} (\bibinfo{year}{2020}).

\bibitem[{\citenamefont{Bao et~al.}(2022)\citenamefont{Bao, Cao, Zheng, and
  Hou}}]{bao2022designing}
\bibinfo{author}{\bibfnamefont{R.}~\bibnamefont{Bao}},
  \bibinfo{author}{\bibfnamefont{Z.}~\bibnamefont{Cao}},
  \bibinfo{author}{\bibfnamefont{J.}~\bibnamefont{Zheng}}, \bibnamefont{and}
  \bibinfo{author}{\bibfnamefont{Z.}~\bibnamefont{Hou}},
  \bibinfo{journal}{arXiv preprint arXiv:2209.11419}  (\bibinfo{year}{2022}).

\end{thebibliography}

\end{document}